\documentclass[%
 aip,
 amsmath,amssymb,
reprint,%
]{revtex4-1}

\usepackage{graphicx}
\usepackage{dcolumn}
\usepackage{bm}

\usepackage[utf8]{inputenc}
\usepackage[T1]{fontenc}
\usepackage{mathptmx}
\usepackage{etoolbox}

\usepackage{xcolor}
\usepackage{booktabs}
\usepackage{siunitx}
\usepackage{braket}
\usepackage{hyperref}
\usepackage{cleveref}
\usepackage{bbold}
\usepackage{ulem}

\definecolor{myblue}{RGB}{68,114,196}

\makeatletter
\def\@email#1#2{%
 \endgroup
 \patchcmd{\titleblock@produce}
  {\frontmatter@RRAPformat}
  {\frontmatter@RRAPformat{\produce@RRAP{*#1\href{mailto:#2}{#2}}}\frontmatter@RRAPformat}
  {}{}
}%
\makeatother
\begin{document}

\preprint{AIP/123-QED}
\sffamily
\title[Unusual Stokes shift and vibronic symmetry]{On the unusual Stokes shift in the smallest PPE dendrimer building block:\\Role of the vibronic symmetry on the band origin?}
\author{Joachim Galiana}
\author{Benjamin Lasorne}%
\affiliation{ 
ICGM, Univ Montpellier, CNRS, ENSCM, Montpellier, France
}
\email{benjamin.lasorne@umontpellier.fr}

\date{\today}
\normalem
\begin{abstract}
1,3-bis(phenylethynyl)benzene is the primary chromophore of light-harvesting polyphenylene ethynylene (PPE) dendrimers. It is experimentally known to share the same absorption spectrum as its pair of diphenylacetylene (aka. tolane) meta-substituted branches, yet exhibits an unusual Stokes shift of about 2000 cm$^{-1}$ with respect to its band origin (corresponding to the loss of one vibrational quantum within the antisymmetric acetylenic stretching) in its emission spectrum. We suggest in the present work the unusual but plausible involvement of molecular symmetry selection rules in a situation where the Born-Oppenheimer approximation is far to be valid. Our hypothesis is comforted with quantum dynamics (MCTDH) simulations of absorption and emission UV-visible spectra based on quantum chemistry (TD-DFT) data and a diabatic vibronic coupling Hamiltonian model.

\noindent
\textbf{Keywords:} light-harvesting molecules, PPE dendrimers, UV-vis spectroscopy, Stokes shift, nonadiabatic quantum dynamics, vibronic coupling, conical intersections
\end{abstract}
\maketitle
\section{Introduction}
\label{sec:introduction}
Light-harvesting dendrimers have drawn much interest in many fields of applications for the past few decades, such as in the design of new photovoltaic technologies or photosynthetic processes.\cite{balzani_harvesting_1995,gilat_light_1999,adronov_light-harvesting_2000,balzani_light-harvesting_2003}
Phenylacetylene dendrimers (covalent tree-like macromolecules) are among the most promising $\pi$-conjugated organic systems for harvesting and amplifying the energy of sunlight, based on simple polyphenylene ethynylenes (PPEs) building blocks connected via meta-substitution at phenylene nodes.
Such building blocks are also of interest on their own, as some of their spectroscopic properties are still challenging to understand and interpret.

One of the best known phenylacetylene dendrimer is the nano-star\cite{xu_design_1994}, for which the building blocks are PPEs with different lengths and substitutions schemes (para- or meta-substitution).
The connectivity of the building blocks within the nano-star is such that the structure is two-dimensional, with a sequential threefold branched graph (two donors on each acceptor).
Experimental and theoretical studies so far have suggested two main results. 
On the one hand, the absorption spectra of the building blocks of PPE dendrimers are similar and additive, with mainly local electronic excitations (LE).\cite{kopelman_spectroscopic_1997,swallen_dendrimer_1999,swallen_correlated_2000,rana_steady-state_2001,wu_exciton_2006,palma_electronic_2010} 
On the other hand, the building blocks are arranged such that a local excitation at the periphery of the nano-star is transferred to its center thanks to an intramolecular electronic excitation energy gradient from the external chromophores to the energy-releasing core.\cite{devadoss_energy_1996,shortreed_directed_1997,kleiman_ultrafast_2001,melinger_optical_2002,fernandez-alberti_nonadiabatic_2009,fernandez-alberti_shishiodoshi_2012,fernandez-alberti_non-adiabatic_2016,nelson_electronic_2017}

\begin{figure}[htp]
    \centering
    \includegraphics[width=0.70\linewidth]{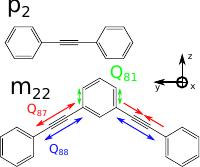}
    \caption{Diphenylacetylene p2 and 1,3-bis(phenylethynyl)benzene m22 molecules. A scheme of the displacements associated to normal modes of vibration 81, 87, and 88 is given for m22.
    The Mulliken axis convention used here is such that in the C\textsubscript{2v} symmetry point group, the normal mode of vibration 87 is of in-plane $y$-type symmetry and belongs to B\textsubscript{2}.}
    \label{fig:p2_m22}
\end{figure}

Among these building blocks, the 1,3-bis(phenylethynyl)benzene (\cref{fig:p2_m22}, bottom, called m22 in the following) shows interesting steady-state spectroscopic properties.
Alike other building blocks based on diphenylacetylene (\cref{fig:p2_m22}, top, called p2 in the following; also known as tolane), it has an absorption spectrum similar in its vibronic structure to the absorption spectrum of diphenylacetylene, only with a different intensity.
This peculiarity was assigned to the additivity of local single excitations on p2 branches in the building blocks, and further rationalized with a pseudo-fragmentation scheme for m22.\cite{thompson_variable_2004,ho_diabatic_2019}
In addition, a minimum energy conical intersection (MECI) was characterized between the first two singlet electronic excited states.\cite{ho_diabatic_2019}
The nature of the coupled electronic singlet excited states was further studied, with two quasi-diabatic representations: one with a locally excited state on each of the p2 branch (right or left), the other with delocalized excited states with a well-defined symmetry: B\textsubscript{2} (right minus left) or A\textsubscript{1} (right plus left).

However, whereas the absorption spectra are almost identical for m22 and p2, the emission spectrum of m22 is completely different from the emission spectrum of p2 experimentally.
An unusual Stokes shift $\Delta \bar{\nu}\simeq \SI{2000}{\per\centi\meter}$ has been reported for the vibronic spectra of m22 separately by different authors.\cite{thompson_variable_2004,gaab_meta-conjugation_2003,chu_vibronic_2004} 
Let us stress out here that we are not considering the typical textbook Stokes shift between the maxima of bell-shaped absorption and emission spectra, essentially due to the influence of a significant displacement between ground- and excited-state equilibrium geometries on vertical absorption and emission transition energies (hence on maximal Franck-Condon factors $0' \rightarrow v''>0$ \emph{vs.} $0'' \rightarrow v'>0$), but rather focusing on the relative spectral shift of the two band origins ($0' \leftrightarrow 0''$), which occur to coincide indeed and be the most intense for p2 but are strangely shifted in the triangular-shaped emission of m22 and not in its triangular-shaped absorption spectrum at low temperature.

A first hypothesis was suggested by some authors\cite{chu_vibronic_2004} of a missing band origin associated to the $0-0$ transition in both the absorption and emission spectra. 
Such a proposition does not seem satisfactory because the $0-0$ transition is not missing in the absorption spectrum of m22, which is similar to the one of p2 and for well-justified reasons. 
It thus seems more likely to investigate some particular effect influencing the emission spectrum of m22 compared to that of p2. 
This is precisely the objective of the present work, where we propose some potential theoretical explanations based on symmetry, yet cannot provide any definitive answer until further time-resolved experiments are carried out.

Gaab, Thompson and co-workers produced an extensive experimental study for m22-like molecules, and proposed a variable excitonic coupling model for the reproduction and interpretation of the Stokes shift in H-terminated meta-substituted phenylacetylene instead of phenyl-terminated ones\cite{thompson_variable_2004,gaab_meta-conjugation_2003}. 
Their model qualitatively agrees with \textit{ab initio} calculations for the ordering of electronic states of interest in the absorption and emission phenomena for meta-substituted PPEs. 
In their work, the theoretical Stokes shift is reproduced by only looking at vertical transition energies for absorption and emission. 
Once again, such a picture implies some type of usual Stokes shift due to some geometrical displacement between the absorbing and emissive species and/or incomplete vibrational relaxation. 
Further along this line, we could invoke some additional effect due to the relaxed or unrelaxed state of the solvent. 
Yet, there seems to some more fundamental effect due to the nature of the absorbing and emissive species while considering steady-state spectra, expected to exhibit the \textit{a priori} common $0' \leftrightarrow 0''$ band origin connecting via light-matter resonance the long-lived (stationary) eigenstates of the system.

As such, we propose hereby a minimal (three-dimensional; 3D) vibronic coupling model, in the fashion of Cederbaum, Köppel, Cattarius, and co-workers\cite{cederbaum_strong_1977,koppel_multimode_1984,cattarius_all_2001} for the construction of a model vibronic Hamiltonian, able to reproduce the vibrational structure of the electronic absorption and emission spectra for m22 via nonadiabatic quantum dynamics techniques beyond the Born-Oppenheimer approximation. This has never been tried on the present system to the extent of our knowledge; we shall use it to discriminate Stokes and non-Stokes shift contributions to the emission spectrum according to light polarization. Let us also stress that such a strategy based on the spectral response of quantum dynamics has been largely tested on photoabsorption or photoelectron/photoionization spectra, but rarely on photoemission spectra, which somewhat remain uncharted territory in terms of spontaneous \textit{vs.} stimulated emission contributions.

With the vibronic coupling models used in this work, the coupling explicitly depends on the geometry of the molecule, through variations of normal mode coordinates, following the primary idea of a variable excitonic coupling proposed by Gaab and Thompson.
If one may say, note that our interpretation is more the one of a chemist point of view (vibronic coupling) than of a physicist point of view (excitonic coupling). 
Indeed, our vibronic coupling models allow for the interpretation of the absorption and emission spectra in terms of vibronic transitions and vibrational overlaps. 
Because the electronic excited states of interest are strongly coupled, the simulation of absorption and emission spectra is a non-trivial procedure for which quantum dynamics was used to overcome the breakdown of the Born-Oppenheimer approximation. 
The low-dimensionality of the model allows us to use the MCTDH (multiconfiguration time-dependent Hartree) method for the propagation of vibrational wave packets on the coupled electronic excited states, thus paving the way to further high-dimensional simulations and/or other types of calculations including the role of the solvent and/or of soft modes in a more effective way, which are works in progress.

As usual in this context, the vibronic spectra are then obtained using the Fourier transform of the autocorrelation functions obtained via the dynamics. 
For the absorption spectra, the results are in very good agreement in spite of the low-dimensionality of the models. 
However, the emission spectra show at this stage two types of contributions: one that complies very well with the observed Stokes shift of the band origin, $\Delta \bar{\nu}\simeq \SI{2000}{\per\centi\meter}$, and another that does not and merely provides coincidence of both band origins. 
The Stokes-shifted band is the one that corresponds with a coherent flip of light polarization consistent with the odd character of the interstate coupling and could be attributed to an emissive Herzberg-Teller contribution or even perhaps a geometric phase; to the extent of our knowledge, this is not an effect that has been reported in large molecules.
However, this may reflect some symmetry-driven selection rule concerning the conservation of orbital momentum that is rarely mentioned in this context.

In the first part, we recall our main insights into the electronic structure calculations on the most important points of the energy landscape of m22, and we give relevant computational details associated to electronic structure and quantum dynamics calculations for this work.
We also describe the vibronic coupling Hamiltonian models that were used in the quantum dynamics calculations hereby.
Next, we discuss the results of these models fitted on accessible \textit{ab initio} data and describe the vibronic eigenstates produced by the models and their molecular-symmetry/light-polarization peculiarities. 
Finally, computed absorption and emission spectra are provided and discussed, and further compared to their experimental counterparts, before conclusions and outlook are given so as to suggest further investigations.
\section{Theoretical background and methods}
\subsection{Conical Intersections and branching-space vectors}
The lowest two excited states of the energy landscape of m22 have been studied previously. \cite{ho_diabatic_2019}
Their potential energy surfaces (PESs) share a conical intersection (CoIn) seam, which lies locally within an $(N-2)$-dimensional subspace (where $N=102$ is the number of internal coordinates for the molecule).
The local complementary subspace to the seam is denoted the branching space, where the energy difference increases linearly from zero along the two branching-space vectors.
The minimum-energy conical intersection (MECI) of the seam was characterized\cite{ho_diabatic_2019} and will be used as a central point in this work.
For high-symmetry points, the branching-space vectors can be associated to simply defined quantities: the (halved) gradient difference vector (\textbf{GD}) and the derivative coupling vector (\textbf{DC}).
In our case, the point group of the m22 molecule is C\textsubscript{2v}, and the \textbf{GD} and \textbf{DC} expand along A\textsubscript{1} normal modes (totally symmetric) and B\textsubscript{2} normal modes (in-plane, non-totally symmetric), respectively.

The usual Born-Oppenheimer approximation fully breaks down over regions such as CoIn seams, which requires considering both electronic excited states simultaneously together with their non-adiabatic couplings.
The adiabatic (eigenstate) representation of the corresponding electronic Hamiltonian is not practical because of the singularity with respect to the energy difference in the non-adiabatic couplings.
Thus, a quasi-diabatic representation should be preferred, with smoothly varying functions (with respect to the internal coordinates) for the diabatic potentials and the inter-state couplings.
In this work, such quantities are, indeed, smoothly varying functions of the internal coordinates associated to acetylenic and quinoidal normal modes of vibration of m22 (\cref{fig:p2_m22}).
\subsection{Computational details}
\subsubsection*{Quantum chemistry}
\label{sec:details_qm}
All quantum chemistry calculations were performed with the Gaussian16 package (revision A.03)\cite{g16} using DFT (for the ground state) and TD-DFT (for the excited states) at the CAM-B3LYP/6-31+G* level of theory.
Such computational choices were already validated for para-conjugated PPE\cite{adamo_calculations_2013,ho_vibronic_2017} and this same meta-substituted molecule for the study of its absorption spectrum, with good agreement to the experiment.\cite{ho_diabatic_2019}
The defect of TD-DFT for the evaluation of charge transfer (CT) contributions to the excited states is not troublesome in the excited state of interest in this work for m22 (the first two excited states are essentially locally excited (LE) states).

The MECI of interest here was taken from the work of E. K. L. Ho \emph{et al.}\cite{ho_diabatic_2019}, and has been optimized in the C\textsubscript{2v} subspace of the molecular geometries.
In the current TD-DFT implementation of the Gaussian package, analytic non-adiabatic couplings (NAC) between excited states and thus the derivative-coupling vector (DC) are not routinely available.
For this reason, the branching-space vectors are evaluated at the MECI using the energy-based and wave-function free method of Gonon \emph{et al.}, along with their lengths, through the diagonalization of the Hessian of the squared energy difference.\cite{gonon_applicability_2017}
The main contribution in \textbf{GD} comes from the quinoidal mode 81 (43\% of the GD), the main contribution in \textbf{DC} comes from the acetylenic mode 87 (73\% of the DC).
Because mode 87 mainly is the antisymmetric elongation of left and right acetylene bonds (B\textsubscript{2}), its symmetric counterpart, the acetylenic mode 88 (A\textsubscript{1}), is added to the description; it also participates in the \textbf{GD} (10\% of the GD). 
Both A\textsubscript{1} modes 81 and 88 are important for the description of the region between the Franck-Condon point (FC) and the MECI. Indeed, both modes have non-negligible contributions to the gradients at the Franck-Condon (FC) point and are associated to important shifts between the FC point and the MECI ($d_{81} = -7.36 \sqrt{m_e}a_0$ and $d_{88}=6.78 \sqrt{m_e}a_0$).

Rigid scans of the PESs of S\textsubscript{0}, S\textsubscript{1}, and S\textsubscript{2} were obtained starting from the MECI and following the normal modes directions computed at the minimum of the electronic ground state.
The grid points are described in \cref{tab:grid}, and the displacement associated to the chosen normal modes of vibration are schematized in \cref{fig:p2_m22}.
\begin{table*}[!hbt]
\caption{Frequency calculation results, symmetry, and generated points along modes 81, 87, and 88 originated from the MECI.}
\label{tab:grid}
\centering
\begin{tabular}{rllll}
  \toprule
  Mode S\textsubscript{0} (S\textsubscript{1}) & Frequency (cm\textsuperscript{-1}) & Reduced mass (AMU) & Symmetry & Grid ($\sqrt{m_e}a_0$)\\
  \midrule
  81 & 1656 & 5.465 & A\textsubscript{1} & 15 points [-18,18]\\
  87 & 2367 (2184) & 11.998 (11.979) & B\textsubscript{2} & 15 points [-18,18]\\
  88 & 2367 (2353) & 11.998 (11.997) & A\textsubscript{1} & 15 points [-18,18]\\
  \bottomrule
\end{tabular}
\end{table*}
B\textsubscript{2} displacements break C\textsubscript{2v} geometries to C\textsubscript{s} geometries (exploring the minima of the S\textsubscript{1} PES) whereas A\textsubscript{1} displacements conserve the C\textsubscript{2v} symmetry of the molecule allowing for an assignation of the diabatic PES using unambiguous symmetry arguments.

\subsubsection*{Quantum dynamics}
\label{sec:details_qd}
The Quantics package was used for quantum dynamics calculations\cite{beck_multiconfiguration_2000,worth_quantics_2020}, mainly for the multiconfiguration time-dependent Hartree (MCTDH) wave packet propagation method. 
Details concerning the single-particule functions (SPFs) and the primitive basis set are given in \cref{tab:quantics_parameters}.
The multi-set formulation of the MCTDH ansatz was used, with different sets of SPFs (8 per state). 
The primitive basis set consists in harmonic oscillator wave functions for the three degrees of freedom.
Two types of calculations were done: relaxations and propagations.
Relaxations are used for computing vibronic excited eigenstates within our three-mode two-state model.
As a result, vibronic states expand in a basis of two electronic wave functions, which are here the delocalized B\textsubscript{2 and A\textsubscript{1} diabatic states}.
Propagations are used for computing autocorrelation functions and associated spectra (for absorption and emission).
The autocorrelation function is computed as:
\begin{equation}
C(t)=\Braket{\Psi(0)|\Psi(t)}=\Braket{\Psi(t/2)^*|\Psi(t/2)}
\end{equation}
and the intensity of the spectra are:
\begin{equation}
I(\omega)\propto \int_{-\infty}^{+\infty}C(t)\exp{(i\omega t)}\mathrm{d}t.
\end{equation}
Because the propagation time is finite, the autocorrelation function was multiplied as usual by a decaying function
\begin{equation}
    g(t)=\cos^n(\pi t/2T),
\end{equation}
where $T$ is the propagation time and $n=1$ in our calculations.
Additionally, the autocorrelation function was multiplied by a Gaussian function characterized by the damping time $\tau = (2\sqrt{2\log{2}})/(\Delta \omega)$ to simulate the experimental full width at half-maximum (FWHF), $\Delta \omega$.
\begin{table*}[!hbt]
\caption{Single-particle functions and primitive basis parameters for the relaxation and propagation of nuclear wave packets. Reduced masses for the HO primitive basis are 1 because the coordinates are mass-weighted.}
\label{tab:quantics_parameters}
\centering
\begin{tabular}{cccccc}
  \toprule
Coordinates & Basis Type & Size of the Basis & Eq. Position & Frequency & Reduced Mass\\
  \midrule
Q\textsubscript{87} & HO & 15 & 0.0 & 2365 cm\textsuperscript{-1} & 1.0\\
Q\textsubscript{81} & HO & 15 & 0.0 & 1655 cm\textsuperscript{-1} & 1.0\\
Q\textsubscript{88} & HO & 15 & 0.0 & 2365 cm\textsuperscript{-1} & 1.0\\
  \midrule
Coordinates & SPFs per state &  &  &  & \\
  \midrule
Q\textsubscript{87} & multi-set 8,8,8 &  &  &  & \\
Q\textsubscript{81} & multi-set 8,8,8 &  &  &  & \\
Q\textsubscript{88} & multi-set 8,8,8 &  &  &  & \\
  \bottomrule
\end{tabular}
\end{table*}
For absorption spectra calculations (ground state to excited states) and for vibronic excited states relaxation, the initial wave packet simply consists in the zero-th order harmonic oscillator wave function with corresponding frequencies for the three coordinates 81, 87, and 88.
For emission spectra calculations (excited states to ground state), the initial wave functions are the vibronic excited states obtained from the relaxations in the electronic excited states.
The different contributions to the emission spectra are discussed in the results section.

\subsection{The Quadratic and Linear vibronic coupling Hamiltonian models (QVC, LVC)}
As mentioned in \cref{sec:introduction}, three modes are chosen for modelling the potential energy surfaces.
They have a specific symmetry, which makes more natural the definition of a diabatic vibronic coupling Hamiltonian model.\cite{cederbaum_strong_1977,koppel_multimode_1984,cattarius_all_2001}
Two levels of description are considered here: a quadratic vibronic coupling (QVC) model, relying on a second-order expansion of the diabatic potential energy surfaces, or a linear vibronic coupling (LVC) model, tuning and coupling the two diabatic potential energy surfaces through linear terms only.

Using a reduced-dimensionality model (here, 3D: 81, 87, 88) implies to make a choice regarding the other $N-3=99$ values of the implicit (frozen) coordinates. 
The MECI was chosen as our “reference” point (where the model coincides exactly with the \textit{ab initio} data pointwise per definition and locally to first order per construction). 
The \textit{ab initio} energies of the S\textsubscript{1} and S\textsubscript{2} PESs (for further fitting our LVC/QVC models) were sampled over displacements along modes 81, 87, and 88 from this “reference” point. 
Our motivation here was to find an optimally balanced description of internal vibrational relaxation effects for describing both the emission process and the effect of vibronic couplings on it. 
On the other hand, our system of internal coordinates is originated from the S\textsubscript{0} minimum (the so-called FC point), $\boldsymbol{Q} = 0$, hence, the centre of the wavepacket after sudden absorption for a realistic harmonic description of the initial S\textsubscript{0} state. 
In 3D, it is in fact only an “apparent FC point” (see \cref{tab:critical_points_pes}): same coordinates as the MECI except for the three active coordinates that are set to zero. 
This slightly deteriorates the description of the absorption in principle, but only moderately in practice (the S\textsubscript{1} and S\textsubscript{2} transition energies are similar in values and preserve the \textit{ab initio} ordering with respect to B\textsubscript{2} and A\textsubscript{1} symmetries of both electronic wavefunctions).
The choice of the “apparent FC point” for the expansion of the Hamiltonian implies that the $\kappa_i^{(k)}$-parameters in the model correspond to the energy gradients (in 3D) at this point; their values are similar to the \textit{ab initio} energy gradients at the true FC point (fully optimized geometry of the minimum in S\textsubscript{0}).

In the following, electronic state 1 (notation superscript (1)) corresponds to the singlet excited A\textsubscript{1} state and electronic state 2 (notation superscript (2)) corresponds to the singlet excited B\textsubscript{2} state.
The diabatic representation of the Hamiltonian in the QVC model for the three modes of the m22 molecule reads:
\begin{widetext}
\begin{equation}
\label{eq:general}
\begin{aligned}
\hat{H}^{\text{dia}}&=
\left(
\hat{T}_{\text{nu}}(\boldsymbol{Q})
\right)\mathbb{1}_2
+
\begin{bmatrix}
E^{(1)}(\boldsymbol{Q}=0) & 0 \\
0 & E^{(2)}(\boldsymbol{Q}=0)
\end{bmatrix}\\
&+
\frac{1}{2}
\begin{bmatrix}
k_{81}^{(1)}Q_{81}^2 + k_{88}^{(1)}Q_{88}^2 + k_{87}^{(1)}Q_{87}^2 & 0 \\
0 & k_{81}^{(2)}Q_{81}^2 + k_{88}^{(2)}Q_{88}^2 + k_{87}^{(2)}Q_{87}^2 &
\end{bmatrix}\\
&+
\begin{bmatrix}
\kappa_{81}^{(1)}Q_{81} + \kappa_{88}^{(1)}Q_{88} & 0 \\
0 & \kappa_{81}^{(2)}Q_{81} + \kappa_{88}^{(2)}Q_{88}
\end{bmatrix}\\
&+
\begin{bmatrix}
0 & \lambda_{87} Q_{87} \\
\lambda_{87} Q_{87} & 0
\end{bmatrix}
+
\begin{bmatrix}
\gamma_{81,88}^{(1)}Q_{81}Q_{88}& 0 \\
0 & \gamma_{81,88}^{(2)}Q_{81}Q_{88}
\end{bmatrix}\\
& +
\begin{bmatrix}
0 & \mu_{87,81}Q_{87}Q_{81} + \mu_{87,88}Q_{87}Q_{88} \\
\mu_{87,81}Q_{87}Q_{81} + \mu_{87,88}Q_{87}Q_{88} & 0 \\
\end{bmatrix}.
\end{aligned}
\end{equation}
\end{widetext}
In this expression:
\begin{enumerate}
\item the first matrix represents the Kinetic Energy Operator (KEO), identical for each electronic state;
\item the second matrix represents the adiabatic energies (vertical excitation energies) at the Franck-Condon point (which is the reference point for the expansion);
\item the third matrix represents the distortion matrix in the framework of the Franck-Condon principle and of the harmonic approximation for vibronic transitions;
\item the fourth matrix represents the first-order description of the MECI with respect to the tuning modes;
\item the fifth matrix represents the first-order description of the MECI with respect to the coupling mode;
\item the sixth matrix represents the second-order description of the MECI with respect to the tuning modes, corresponding to the \emph{primary Duschinsky} transformation or mode mixing between modes of same symmetry;
\item the seventh matrix represents the second-order description of the MECI with respect to the coupling mode, corresponding to the \emph{secondary Duschinsky} transformation or mode mixing between modes of different symmetry.
\end{enumerate}
In this context, we call \emph{primary Duschinsky} parameters the second-order coupling parameters between normal modes of identical symmetry (81 with 88) and \emph{secondary Duschinsky} parameters the second order coupling parameters between normal modes of different symmetry (87 with 81 or 88). 
The \textit{ab initio} Duschinsky coefficients obtained for the correlation between the first electronic excited state and the electronic ground state are given in the \cref{sec:duschinsky}, along with the shift vector. 

Such a full-QVC 3D model should grasp most of the physics of the problem, with a total of 17 parameters.
However, as will be pointed out in the results section, further simplifications can be made without significant loss of information.
Hence, in the parent LVC model, all bilinear terms are neglected, such that the Hamiltonian reads:
\begin{widetext}
\begin{equation}
\label{eq:general_2}
\begin{aligned}
\hat{H}^{\text{dia}}&=
\left(
\hat{T}_{\text{nu}}(\boldsymbol{Q})
\right)\mathbb{1}_2
+
\begin{bmatrix}
E^{(1)}(\boldsymbol{Q}=0) & 0 \\
0 & E^{(2)}(\boldsymbol{Q}=0)
\end{bmatrix}\\
&+
\frac{1}{2}
\begin{bmatrix}
k_{81}^{(1)}Q_{81}^2 + k_{88}^{(1)}Q_{88}^2 + k_{87}^{(1)}Q_{87}^2 & 0 \\
0 & k_{81}^{(2)}Q_{81}^2 + k_{88}^{(2)}Q_{88}^2 + k_{87}^{(2)}Q_{87}^2 &
\end{bmatrix}\\
&+
\begin{bmatrix}
\kappa_{81}^{(1)}Q_{81} + \kappa_{88}^{(1)}Q_{88} & 0 \\
0 & \kappa_{81}^{(2)}Q_{81} + \kappa_{88}^{(2)}Q_{88}
\end{bmatrix}\\
&+
\begin{bmatrix}
0 & \lambda_{87} Q_{87} \\
\lambda_{87} Q_{87} & 0
\end{bmatrix}
\end{aligned}
\end{equation}
\end{widetext}
In both expansions of the model, diagonal terms are the diabatic potential energies associated to delocalized electronic states, with specified symmetry.
Thus, the electronic states \textsuperscript{1}A\textsubscript{1} and \textsuperscript{1}B\textsubscript{2} are coupled through the $\text{B\textsubscript{2}}=\text{A\textsubscript{1}}\bigotimes \text{B\textsubscript{2}}$ normal mode of vibration, 87.
The eigenvalues of the model Hamiltonian are to be compared with the adiabatic energies of the first two electronic excited states computed with \textit{ab initio} quantum chemistry calculations.
\subsection{Evaluating the parameters of the LVC and QVC models}
\label{sec:fitting_LVC_QVC}
In the present L/QVC models, the diabatic states 1 and 2 have distinct symmetries when $Q_{87}=0$, that is in the C\textsubscript{2v} subspace of the molecular geometries.
This subspace is partly described by the profiles of 81 and 88, and energies from quantum chemistry calculation can be assigned to either the A\textsubscript{1} or the B\textsubscript{2} diabatic states (which fully identify to either adiabatic states S\textsubscript{1} or S\textsubscript{2} depending on the side of the MECI the calculation was made).
Thus, all parameters except for the one related to the mode 87 can be optimized by fitting directly the diabatic potential energies onto the symmetry-assigned potential energy surfaces from \textit{ab initio} calculations, such that $\{H_{11}(Q_{81},Q_{88}),H_{22}(Q_{81},Q_{88})\}$ are simultaneously fitted to the data $\{E_{\text{A\textsubscript{1}}},E_{\text{B\textsubscript{2}}}\}$ when $Q_{87}=0$. The fitting procedure uses the minimization of the squared-deviation functionals defined as:
\begin{equation}
  \label{eq:first_fit}
  \begin{aligned}
  &L(E^{(1)},E^{(2)},\boldsymbol{\kappa},\boldsymbol{\gamma},\boldsymbol{k}_{81},\boldsymbol{k}_{88})=\\
  &\sum_{n, Q_{87}(n)=0}\left(H_{11}[E^{(1)},\boldsymbol{\kappa}^{(1)},\boldsymbol{\gamma}^{(1)}](\boldsymbol{Q}(n))-E_{\text{A\textsubscript{1}}}(n)\right)^2\\
  +&\sum_{n, Q_{87}(n)=0}\left(H_{22}[E^{(2)},\boldsymbol{\kappa}^{(2)},\boldsymbol{\gamma}^{(2)}](\boldsymbol{Q}(n))-E_{\text{B\textsubscript{2}}}(n)\right)^2.
  \end{aligned}
\end{equation}
Then, taking the resulting parameters for granted, the eigenvalues of the full LVC and QVC models are fitted to the data from maps (87,81) and (87,88) so as to optimize the parameters related to mode 87.
Again, the parameters are obtained through a least-square fitting procedure minimizing the function: 
\begin{equation}
  \label{eq:second_fit}
  \begin{aligned}
  &L(\lambda_{87},\boldsymbol{\mu},\boldsymbol{k}_{87})=\\
  &\sum_{n}\left(V_{1}[\lambda_{87},\boldsymbol{\mu},\boldsymbol{k}_{87}](\boldsymbol{Q}(n))-E_{\text{S\textsubscript{1}}}(n)\right)^2\\
  + &\sum_{n}\left(V_{2}[\lambda_{87},\boldsymbol{\mu},\boldsymbol{k}_{87}](\boldsymbol{Q}(n))-E_{\text{S\textsubscript{2}}}(n)\right)^2,
  \end{aligned}
\end{equation}
where $V_1(\boldsymbol{Q})$ and $V_2(\boldsymbol{Q})$ are the eigenvalues of the model Hamiltonians. 

For the LVC model, only separate profiles along 81, 87, and 88 are required, whereas the QVC model requires 2D-surfaces (87,81), (87,88), and (81,88) to optimize the second-order parameters.
Both yield parameters that allow the reconstruction of the PESs in the 3D-space but the LVC model does not account for full correlation between the coordinates, which will be discussed in the following section.
The fitting procedure with the determination of the L/QVC parameters was done using the functions of the \texttt{SciPy} library in \texttt{Python3.4}.
\section{Results and discussion}
\subsection{\textit{Ab initio} calculations and LVC parameters}
In this section we focus on the use of the LVC model only, for which the parameters are given in \cref{tab:parameters_LVC}.
Note that although the fitting procedure was carried out from data in atomic units and corresponding to  mass-weighted coordinates, the parameters are given in understandable equivalent quantities, such as frequencies (actually, wavenumbers) and geometrical shifts instead of curvatures and gradients, respectively. 
\begin{table}[!hbt]
\caption{LVC parameters obtained upon fitting \textit{ab initio} calculations. Parameters associated to the first-order expansion ($\lambda_{i}$ or $\kappa_{i}$) are given by the associated characteristic shift $d_{i}=\pm\frac{\lambda_{i}}{k_{i}}\text{ or }-\frac{\kappa_{i}}{k_{i}}^{(k)}$ in the mass-weighted framework for displacement along the normal modes directions.
Superscripts (1) and (2) refer to electronic excited states A\textsubscript{1 and B\textsubscript{2}, respectively, coincident with S\textsubscript{2} and S\textsubscript{1} at the FC point.}}
\label{tab:parameters_LVC}
\centering
\begin{tabular}{cccc}
\toprule
Parameter & Equivalent & Value & Unit\\
  \midrule
  $E^{(1)}$ & -- & 4.405 & \si{\electronvolt}\\
  $E^{(2)}$ & -- & 4.380 & \si{\electronvolt}\\
  \midrule
  $\kappa_{81}^{(1)}$ & ${d_{81}^{(1)}}$  & 7.900 & $a_{0}\sqrt{m_{e}}$ \\
  $\kappa_{81}^{(2)}$ & ${d_{81}^{(2)}}$ & 3.094 & $a_{0}\sqrt{m_{e}}$ \\
  $\kappa_{88}^{(1)}$ & ${d_{88}^{(1)}}$ & -7.205 & $a_{0}\sqrt{m_{e}}$ \\
  $\kappa_{88}^{(2)}$ & ${d_{88}^{(2)}}$ & -8.274 & $a_{0}\sqrt{m_{e}}$ \\
  $\lambda_{87}$ & $d_{87}$ & 7.417 & $a_{0}\sqrt{m_{e}}$ \\
  \midrule
  $k_{81}^{(1)}$ & $\omega_{81}^{(1)}$& 1515 & \si{\per\centi\meter}\\
  $k_{81}^{(2)}$ & $\omega_{81}^{(2)}$& 1552 & \si{\per\centi\meter}\\
  $k_{88}^{(1)}$ & $\omega_{88}^{(1)}$& 2274 & \si{\per\centi\meter}\\
  $k_{88}^{(2)}$ & $\omega_{88}^{(2)}$& 2281 & \si{\per\centi\meter}\\
  $k_{87}^{(1)}$ & $\omega_{87}^{(1)}$& 2200 & \si{\per\centi\meter}\\
  $k_{87}^{(2)}$ & $\omega_{87}^{(2)}$& 2201 & \si{\per\centi\meter}\\
  \bottomrule
\end{tabular}
\end{table}

\begin{figure}[!htb]
    \centering
    \includegraphics[height=0.7\textheight]{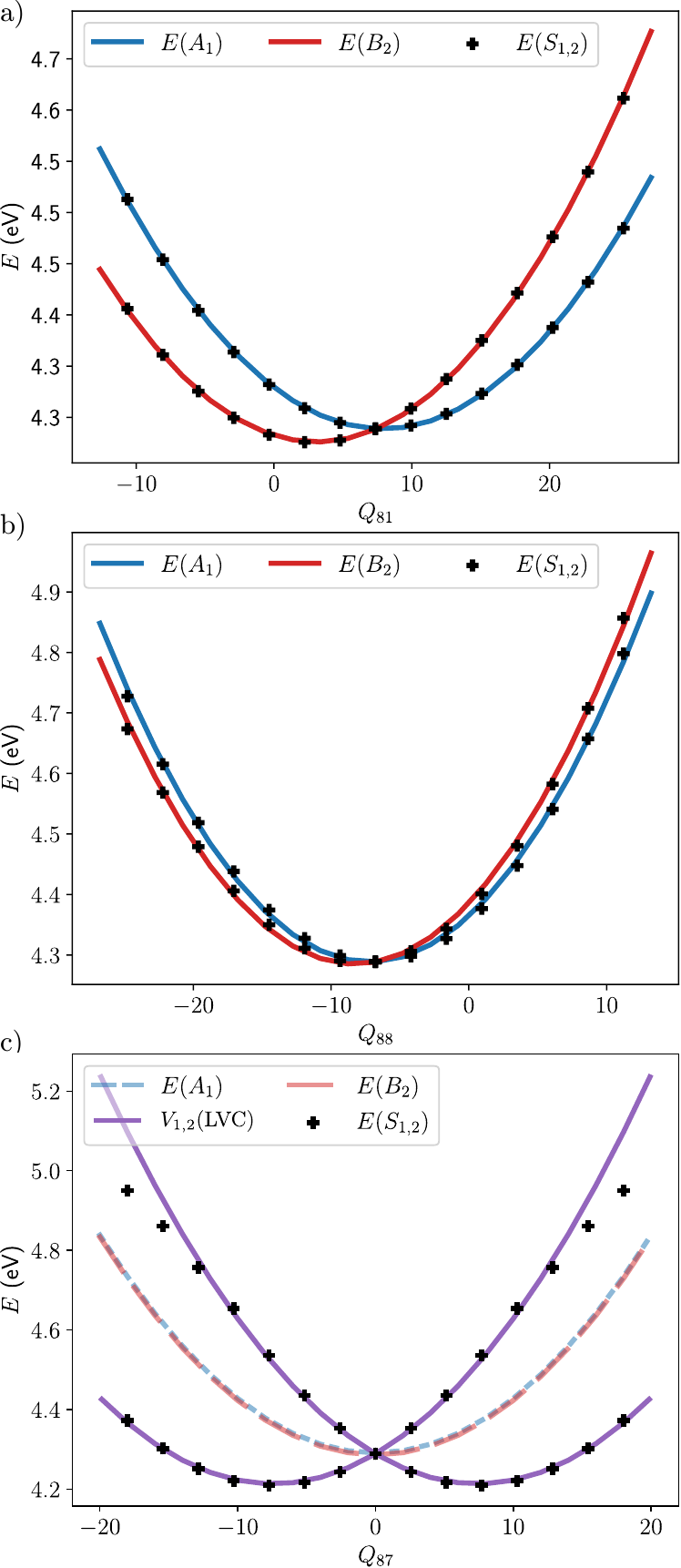}
    \caption{Adiabatic energies from quantum chemistry calculations (symbol $\boldsymbol{+}$) and diabatic potential energies from the delocalized model (blue and red dashed lines) along the 81\textsuperscript{th} (a), 88\textsuperscript{th} (b) and 87\textsuperscript{th} (c) normal modes of vibration. 
    Adiabatic energies (eigenvalues) from the LVC Hamiltonian model are also given for mode 87 (c, plain purple lines).
    For unspecified coordinates, the values are those at the MECI.
    All coordinates are mass-weighted and given in atomic units.}
    \label{fig:profiles_all}
\end{figure}

\Cref{fig:profiles_all} shows the adiabatic energies of states S\textsubscript{1} and S\textsubscript{2} from \textit{ab initio} calculations and the potential energy profiles computed with the latter optimized parameters.
The profiles for modes 81 and 88 are analogous.
Displacements along these A\textsubscript{1} normal modes lead to neither even nor odd potential energy surfaces with respect to the associated A\textsubscript{1} coordinates.
These displacements preserve the C\textsubscript{2v} geometry and thus the delocalized quasi-diabatic electronic states (based on symmetry labelling) maintain their nature.
Thus, we can assign \textit{ab initio} adiabatic energies ($E(\text{S\textsubscript{1,2}})$ symbol \textbf{+} in \cref{fig:profiles_all}, a and b) directly to the diabatic potentials ($E^{(1)}=E(\text{A\textsubscript{1}})$ and $E^{(2)}=E(\text{B\textsubscript{2}})$) of the LVC model.
Note that displacements along the normal mode 81 lift more efficiently the degeneracy than along 88, which is consistent with the fact that the gradient difference vector \textbf{GD} overlaps more with the direction associated to the normal mode 81 than 88. 
The double-well shape of S\textsubscript{1} profile along mode 87 reflects the effect of the antisymmetrical coupling along the \textbf{DC} between the delocalized diabatic states.
This implies that for mode 87, the eigenvalues of the LVC model Hamiltonian are directly fitted to the adiabatic energies from \textit{ab initio} calculations.

\begin{figure}[!hbt]
\centering
\includegraphics[width=1.0\linewidth]{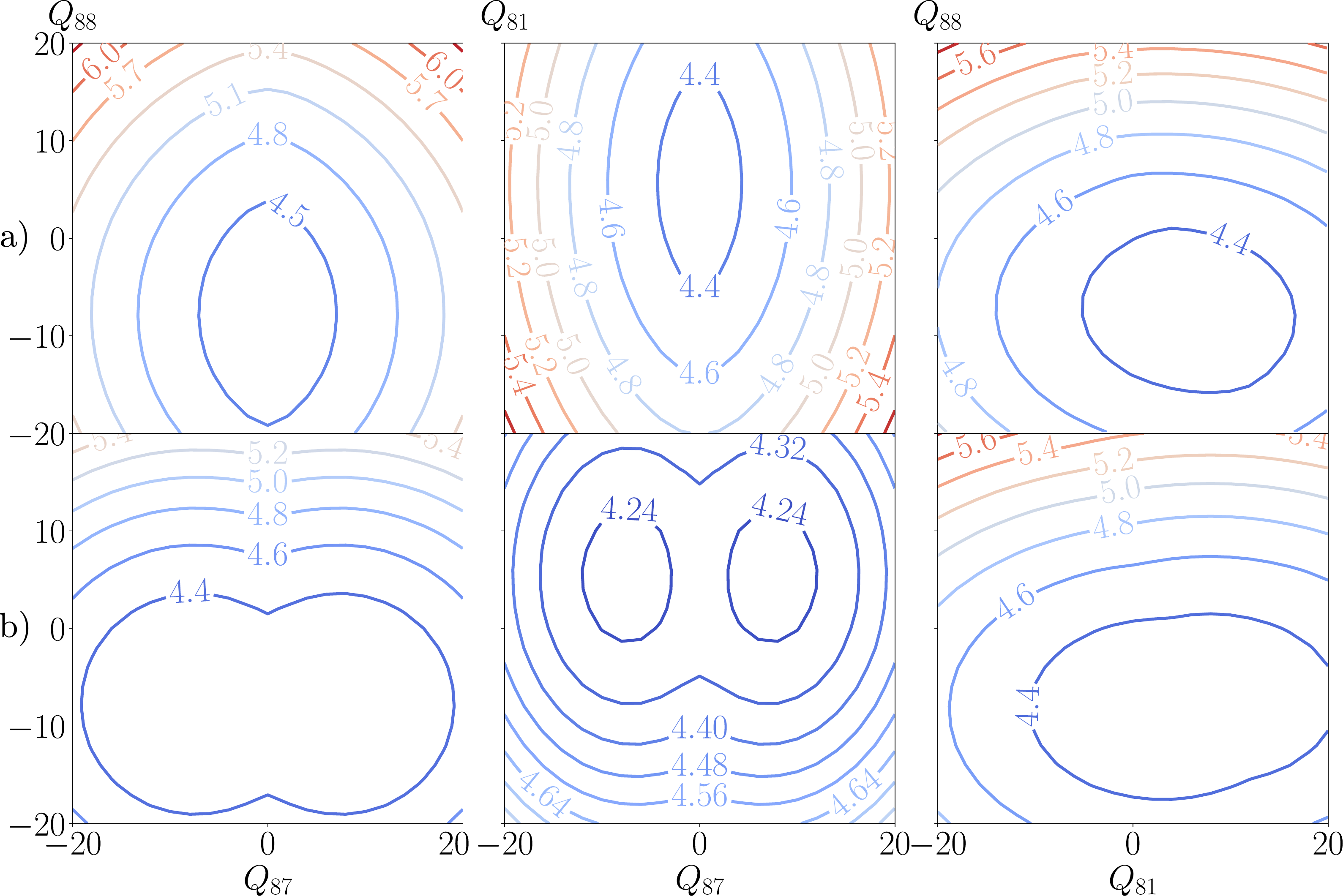}
\caption{LVC potential energy surfaces (eigenvalues of the LVC Hamiltonian model in \si{eV}) of electronic states S\textsubscript{1} (a) and S\textsubscript{2} (b) in the planes (87,81), (87,88), and (81,88) from left to right. 
For unspecified coordinates, the values are those at the MECI.
All coordinates are mass-weighted and given in atomic units, $\sqrt{m_e}a_0$.}
\label{fig:contours_all}
\end{figure}

Finally, these parameters give a three-dimensional picture in the vicinity of the MECI, with exact description of the MECI (starting point of our calculations) and accurate description of the minima in the first electronic excited state. 
In \cref{fig:contours_all}, we give the LVC model PESs of the first two electronic excited states in the planes (87,81), (87,88), and (81,88).
The minima of S\textsubscript{1} are correctly reproduced, with  $E=\SI{4.21}{\electronvolt}$ which is in a good agreement with the \textit{ab initio} calculations ($E=\SI{4.12}{\electronvolt}$). 
The deviation from the \textit{ab initio} optimized minima of S\textsubscript{1} is reasonable considering the following: (i) the starting point for the rigid scans is a MECI and not one of the optimized minima, and (ii) the explored space is only a restricted three-dimensional subspace within all possible molecular geometries. 
The positions of the minima are consistent with the shift vector, given in \cref{sec:duschinsky}, with two equivalent minima with respect to the non-totally symmetric coordinate 87.
The positions are also in good agreement for the totally symmetric coordinates 81 and 88, with the minima being in the positive region of 81 (quinoidal elongation) and negative region of 88 (acetylenic elongation).
The comparison of the critical points energy in the \textit{ab initio} PES and in the parameterized PES is summarized in \cref{tab:critical_points_pes}.
The most important deviation is found for the vertical transition energies and for the minimum of the S\textsubscript{1} state.
\begin{table}[!htb]
    \caption{Energies of the critical points in the \textit{ab initio} PESs (CAM-B3LYP/6-31+G\textsubscript{*} level of theory from Ho \emph{et al.}\cite{ho_diabatic_2019}) and in the parameterized LVC model PESs.
    }
    \label{tab:critical_points_pes}
    \centering
    \begin{tabular}{l|cccccc}
        \toprule
        Critical Point & MinS\textsubscript{1} & TSB\textsubscript{2} & TSA\textsubscript{1} & $E_{S\textsubscript{1}}(\boldsymbol{0})$ & $E_{S\textsubscript{2}}(\boldsymbol{0})$ & MECI \\
        \midrule
         \small{CAM-B3LYP/6-31+G\textsubscript{*}}  & 4.12 & 4.25 & 4.29 & 4.43 & 4.47 & 4.29 \\
         \small{LVC Model} & 4.209 & 4.288 & 4.272 & 4.380 & 4.405 & 4.29 \\
         \bottomrule
    \end{tabular}
\end{table}

As mentioned in \cref{sec:fitting_LVC_QVC}, the LVC model is limited for reproducing the correlation between the coordinates, in particular for taking into account the Duschinsky mixing of the modes between the ground state and the excited states.
The importance of reproducing this correlation with the QVC model is discussed in the next section.
\subsection{Comparison of linear and quadratic VC models}
The LVC and QVC models show almost identical values for the parameters that they share and no significant effects of the extra second-order parameters of the QVC model (see Supplementary Information (SI), tab. SI2). 
Similar figures as the one provided for the LVC model earlier are given in the SI for the QVC model, see fig. SI1.

\begin{figure}[!hbt]
\hspace{-0.025\textwidth}
  \centering
  \includegraphics[width=1\linewidth,angle=0]{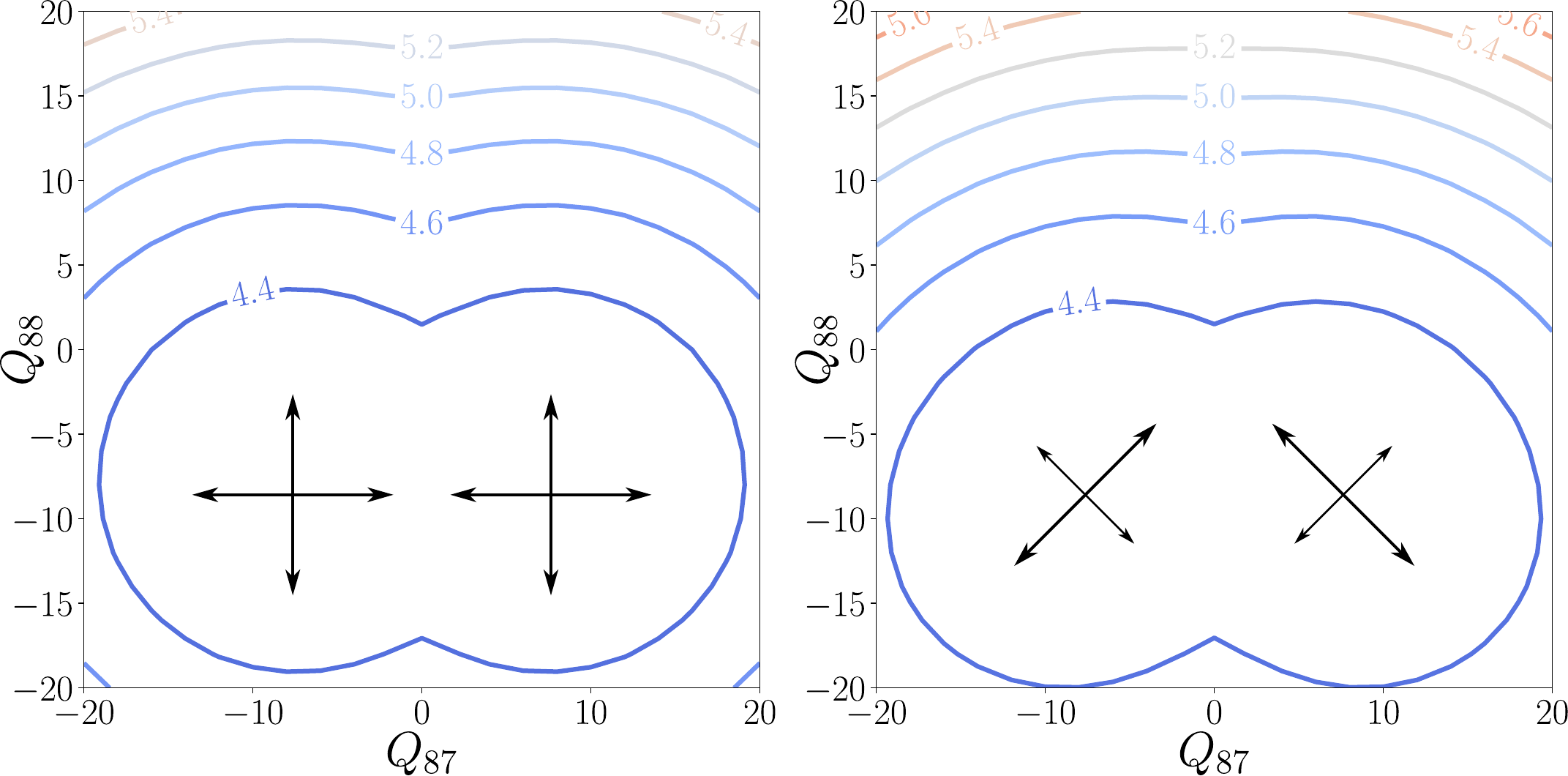}
  \caption{Potential surfaces of the first excited state in the LVC (left) and QVC (right) models with respect to the acetylenic coordinates 87 and 88. Central arrows correspond to the directions of the main axes of the ellipsoids.
  All coordinates are mass-weighted and given in atomic units, $\sqrt{m_e}a_0$.}
  \label{fig:axes}
\end{figure}

The main difference lies in the ellipsoidal shapes of the isoenergy slices through the potential energy surfaces with respect to the direction of the A\textsubscript{1} normal mode 88.
Indeed, with the linear description only (LVC), the ellipsoids are aligned with the directions of the ground state normal modes of vibration (chosen for the generation of the grids), which is in contradiction with the Duschinsky transformation of the normal modes from the electronic ground state to excited states.
This transformation is taken into account in the quadratic description (through bilinear terms), for which the ellipsoids have different main axes, as shown in \cref{fig:axes}.
However, there is no significant effect of this transformation on the physics of the (dimensionally-reduced) system and in particular for the absorption and emission spectra.
This can be explained by the almost equivalent frequencies for modes 87 and 88, which make the rotation of the potential wells chemically motivated (pseudofragmentation, consistent with left or right local modes on each tolane moiety) but physically (or, say, numerically) not significant.

\subsection{Vibronic states in the excited-state landscape}
To calculate and interpret the details of vibronic transitions within UV-visible absorption and emission spectra, we must access the vibronic eigenstates of the molecule and their eigenenergies. 
In the electronic ground state, the vibronic states are simple because the Born-Oppenheimer approximation is essentially valid there. 
We describe those states as simple products of the electronic ground state wave function and the 3D harmonic oscillator wave functions, for which the parameters are the frequencies and reduced masses of the normal modes of vibration 81, 87, and 88 in the electronic ground state.
For the excited states, the description of the vibronic eigenstates is more involved. 
We are in fact facing a situation whereby the Born-Oppenheimer approximation is completely off.
Due to the quasi-degeneracy of the two states of interest, already in the Franck-Condon region, and the presence of an accessible conical intersection in its close vicinity and inducing a strong diabatic mixing, the first two excited adiabatic electronic states are strongly coupled nondiabatically and cannot at all be treated separately for the description of the associated vibronic states. 
In other words, and using the language of spectroscopy, we are in the unusual situation where the first-order Herzberg-Teller effect along mode 87 is clearly not a perturbation.

Within a description beyond Born-Oppenheimer the vibronic states expand as linear combinations of products of electronic and nuclear wave functions. 
We choose the delocalized, symmetry-labelled, diabatic electronic basis set, as described earlier for the LVC Hamiltonian model. 
Viewed as a good approximation of a  crude adiabatic basis for the Herzberg-Teller effect, it presents the great advantage that transition dipoles will not vary much with displacements from C\textsubscript{2v} molecular geometries (such as the Franck-Condon ground-state equilibrium geometry); in particular, A\textsubscript{1} is polarized along $z$ and B\textsubscript{2} along $y$. 
Within this representation, we thus get:
\begin{equation}
\begin{aligned}
\Psi^{\text{(exc)}}_k (\boldsymbol{q}, \boldsymbol{Q})
&=
\chi_{A_1,k} (\boldsymbol{q}) \phi_{A_1}(\boldsymbol{q};\boldsymbol{Q})\\
&+
\chi_{B_2,k} (\boldsymbol{q}) \phi_{B_2}(\boldsymbol{q};\boldsymbol{Q})
\end{aligned}
\end{equation}
where $\boldsymbol{Q}=(Q_{81},Q_{87},Q_{88})$ are the internal coordinates and $\boldsymbol{q}$ are the electronic coordinates.
The electronic states $\phi_{A_1}(\boldsymbol{q};\boldsymbol{Q})$ and $\phi_{B_2}(\boldsymbol{q};\boldsymbol{Q})$  are quasi-diabatic states, which vary moderately and smoothly (ideally not at all to first order) with the internal coordinates $\boldsymbol{Q}$. 
The associated diabatic state populations are simply the integral of the vibrational contribution over the internal coordinates $\boldsymbol{Q}$.
In this work, such vibronic eigenstates are obtained through quantum dynamics calculations on the “coupled” PESs from the LVC Hamiltonian model.  
The calculation consists in the propagation in imaginary time of an initial nuclear wave packet until variational convergence. 
The so-called relaxation method is analogous to variational energy minimisation (parametric optimisation). 
We consider two choices of initial states (they will be the same as those used later on for the absorption spectra simulations). 
They are the vibrational ground state for the electronic ground singlet state (fundamental of the 3D harmonic oscillator for 81, 87, and 88) multiplied by either the diabatic state A\textsubscript{1} or the diabatic state B\textsubscript{2} (coupled electronic excited singlet states). 

\begin{figure}[!htb]
\centering
\includegraphics[width=1\linewidth]{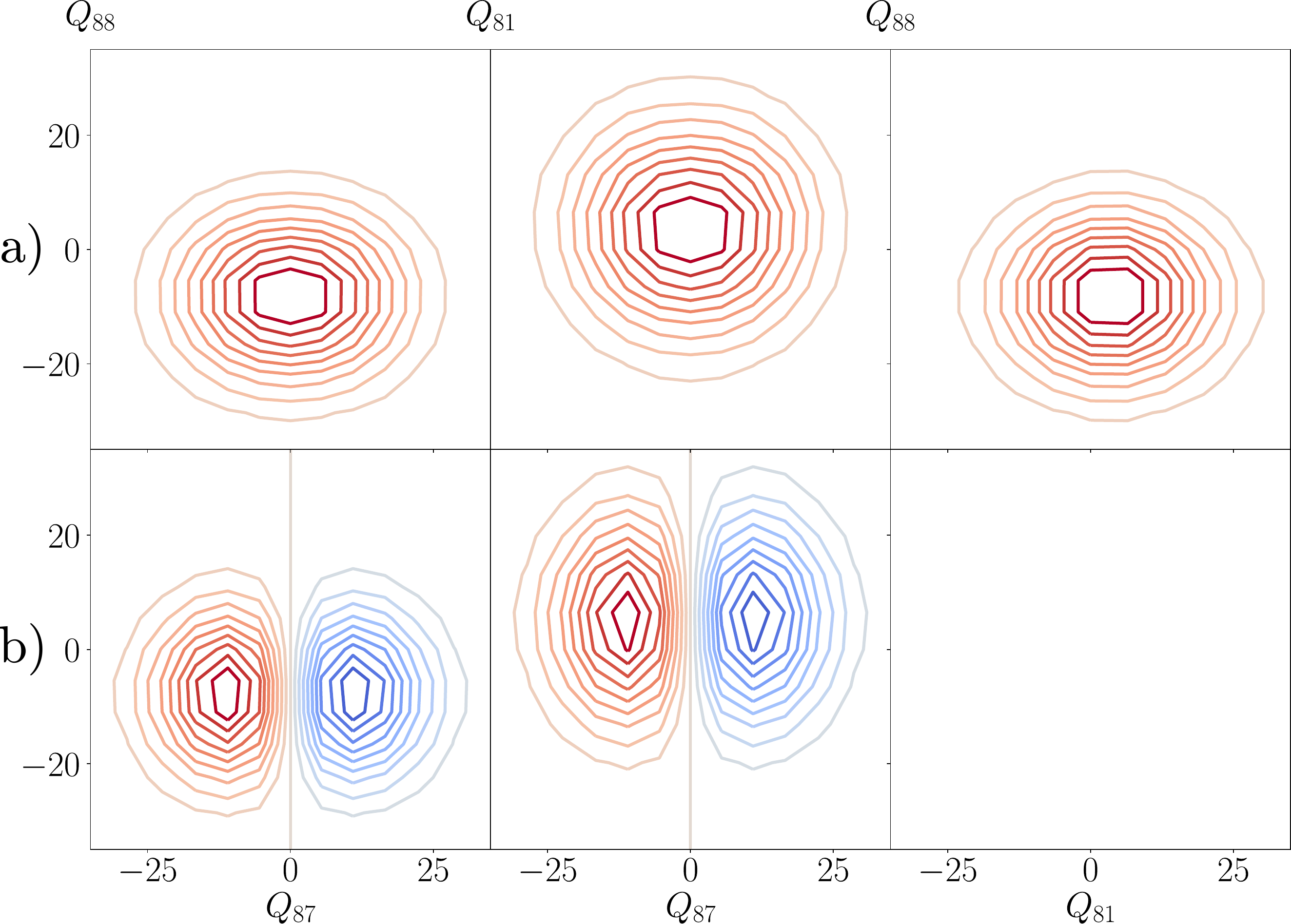}
\caption{Nuclear wave packet contributions in diabatic electronic states B\textsubscript{2} (a) and A\textsubscript{1} (b) in the ground vibronic state of the coupled S\textsubscript{1/S\textsubscript{2} manifold}, in planes (87,81), (87,88), and (81,88) from left to right. 
For unspecified coordinates, the values are those at the MECI.
All coordinates are mass-weighted and given in atomic units, $\sqrt{m_e}a_0$.
The blank figure corresponds to the wave functions having the B\textsubscript{2} coordinate 87 in the nodal plane.
The first-excited vibronic state of the coupled S\textsubscript{1/S\textsubscript{2} manifold} is described in fig. SI3
}
\label{fig:vibronic_states}
\end{figure}

\begin{figure}[!hbt]
\centering
\includegraphics[width=1\linewidth]{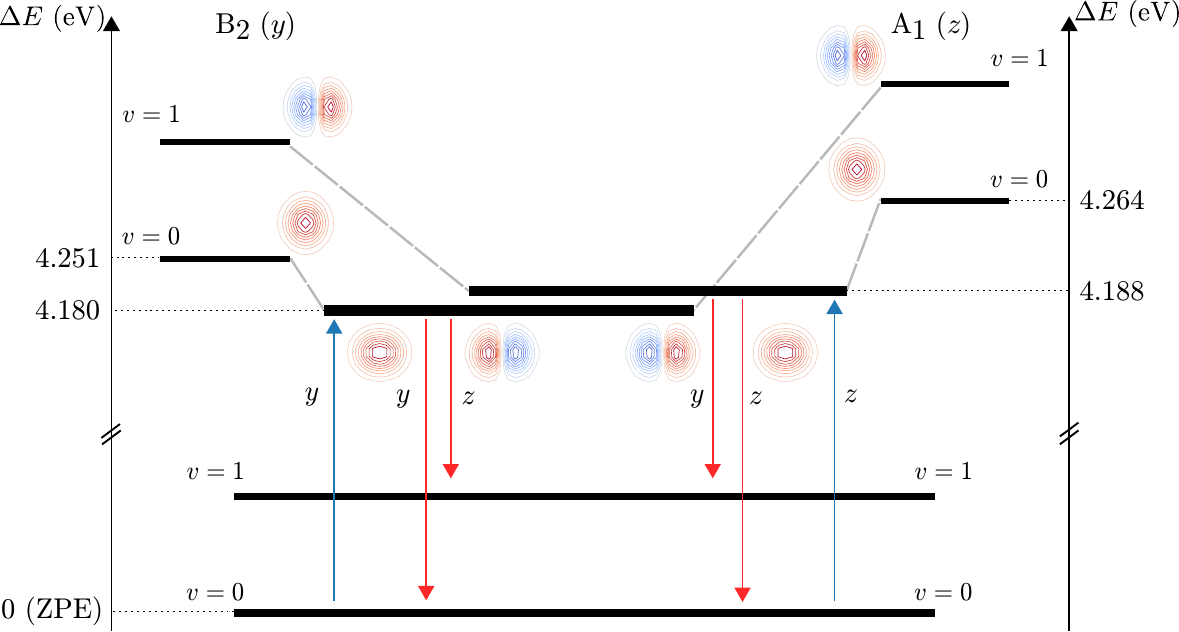}
\caption{Representation of the vibronic eigenstates as interacting two-state(A\textsubscript{1, B\textsubscript{2})-two-body(Q\textsubscript{87}, Q\textsubscript{88}) vibronic wavefunction components}.
The vibrational number $v$ refers to the excitation number of non-totally symmetrical normal mode of vibration 87.
The nuclear wave packets are given in the plane (87,88).}
\label{fig:interaction_diagram}
\end{figure}

After an imaginary relaxation time of about \SI{20}{\femto\second}, which is short and indicates little internal vibrational redistribution, two quasi-degenerate ($\Delta E = \SI{0.0076}{\electronvolt}$) but distinct vibronic states are obtained in the S\textsubscript{1/S\textsubscript{2} manifold}. They could be viewed as a tunnelling pair, symmetric and antisymmetric, in an adiabatic context. 
In the following, the ground vibronic state and the first-excited vibronic state of the coupled S\textsubscript{1}/S\textsubscript{2} manifold are called the first vibronic state and the second vibronic state, respectively, unless otherwise specified.
Their characteristics are collected in \cref{tab:vibronic_states}. 
We notice that both vibronic states are complementary. 
Vibrational contributions of the vibronic states in the  planes (87,81), (87,88), and (81,88) are given (see \cref{fig:vibronic_states} for the lowest, and fig. SI3 for the second vibronic state). 
The first vibronic state is mainly B\textsubscript{2}-populated (80\%) while the second one is mainly A\textsubscript{1}-populated (78\%) which is consistent with the \textit{ab initio} energy and gradient information regarding the two delocalized diabatic states. 
Indeed the minimum of B\textsubscript{2} is lower in energy than the minimum of A\textsubscript{1} in our 3D model, which is in good agreement with the energy ordering of the optimized transition states A\textsubscript{1} and B\textsubscript{2} (see \cref{tab:critical_points_pes}).
These results also indicate that with this model, in each vibronic state, about 80\% of the initial wave packet is preserved in the stationary vibronic states, which represents about 80\% of totally symmetrical vibrational contributions with respect to the three coordinates (see \cref{fig:vibronic_states}, (a)). 
As a result, we notice about 20\% of population transfer to the second diabatic state, not-initially targeted, allowed by the coupling coordinate 87. 
This 20\% transferred population corresponds to an odd shape of the vibrational contributions with respect to the coupling coordinate 87 (see \cref{fig:vibronic_states}, (b)). 
We stress here that the part of the nuclear wave packet that transfers from one diabatic state to another must undergo a change in its parity or symmetry according to the odd behavior of the diabatic interstate coupling function with respect to mode 87. 
This would be termed a geometric (Berry or Longuet-Higgins) phase effect if the adiabatic representation were used instead. 
This change is easily identified in the delocalized and symmetry-labelled diabatic states basis set: the nuclear wave packet turns odd when leaving its initial diabatic state, due to the off-diagonal coupling being odd.
Let us notice that the first two vibronic states are distinct, but similar with only inversion of the roles of the diabatic states A\textsubscript{1} and B\textsubscript{2}. 
\begin{table}[!hbt]
\caption{Vibronic states energy and diabatic populations.
The transition energy is computed by setting the energy reference to the zero-point energy of the 3D model, $E=\SI{0.396}{\electronvolt}$} 
\label{tab:vibronic_states}
\centering
\begin{tabular}{ccccc}
  \toprule
  Vibronic state & Total Energy (eV) & Transition Energy (eV) & $\mathcal{P}_{A_1}$ & $\mathcal{P}_{B_2}$\\
  \midrule
    $\Psi^{\text{(exc)}}_{0}$ & 4.576 & 4.180 & 0.20 & 0.80\\
    $\Psi^{\text{(exc)}}_{1}$ & 4.584 & 4.188 & 0.78 & 0.22\\
  \bottomrule
\end{tabular}
\end{table}
We propose to interpret again the vibronic states looking mostly at their character with respect to the non-totally symmetrical coordinate 87. 
\Cref{fig:interaction_diagram} gives a schematic representation of the vibronic eigenstates in the ground and excited electronic states. 
The pair of vibronic eigenstates in the electronic excited states manifold can be seen as vibronic eigenstates in diabatic states B\textsubscript{2} and A\textsubscript{1} interacting through the non-totally symmetrical coordinate 87.
This representation is analogous to an interaction diagram for electronic orbitals, only with coupled vibrational orbitals here. 
The first two pairs of vibronic eigenstates are the results of four vibrational orbitals interacting and mixing their character. 
As a result, the lower vibronic eigenstate has two vibrational contributions of different parity, odd and even with respect to coordinate 87.

For absorption, the initial wave packet in the ground electronic state overlaps with the even contribution of each of the vibronic eigenstates. 
For emission, two cases arise. 
The initial wave packets have even and odd vibrational contributions, that overlap with the first and second vibronic states in the ground electronic states, respectively. 
Two contributions to the emission spectrum are thus obtained.
One is sharing the $0-0$ band origin with absorption and one is shifted by the vibrational energy of the second vibrational state of the electronic ground state.
Analogous transitions occur for the second vibronic eigenstate. The transition energies to the vibronic eigenstates are ordered as the transition energies of the non-interacting diabatic vibronic eigenstates. 
The latter ordering is consistent with the \textit{ab initio} potential energy minima of the electronic states in the C\textsubscript{2v} sub-space which are transition states of the first excited state in the C\textsubscript{S} space of molecular geometries. 

\subsection{Absorption and emission spectra}
In this section, we describe the procedure for producing absorption and emission spectra (vibronic spectra) using quantum dynamics for our model of the m22 molecule. 
The different contributions to the spectra will be discussed and compared to the low-temperature experimental spectra from Chu and Pang experiments.\cite{chu_vibronic_2004}
Vibronic spectra are computed through Fourier transformation (FT) of the autocorrelation function associated to the propagation of the nuclear wave packet in the electronic excited states (for absorption) or in the electronic ground state (for emission). 
The fact that only the autocorrelation function of the nuclear wave packet is required as a good approximation is helped by the use of the delocalized and symmetry-labelled diabatic basis set for the electronic states. 
Indeed, in this basis set, the transition dipole moments have very little dependence on the nuclear coordinates (Franck-Condon-type approximation):
\begin{equation}
\begin{aligned}
\boldsymbol{\mu}_{\text{A\textsubscript{1}}}(\boldsymbol{Q}) &\simeq \boldsymbol{\mu}_{\text{A\textsubscript{1}}}(\boldsymbol{0}) =\num{-1.8303}\hat{\boldsymbol{z}} (a_0\sqrt{m_e})\\
\boldsymbol{\mu}_{\text{B\textsubscript{2}}}(\boldsymbol{Q}) &\simeq \boldsymbol{\mu}_{\text{B\textsubscript{2}}}(\boldsymbol{0}) =\num{3.9656}\hat{\boldsymbol{y}} (a_0\sqrt{m_e}).
\end{aligned}
\end{equation}
In other words, we can safely neglect Herzberg-Teller terms in the symmetry-adapted delocalized diabatic representation, which, in contrast, would be unusually large with respect to variations in the coordinate 87 in the strongly mixed adiabatic representation, as already pointed out in \cref{sec:introduction}.
As a consequence, the absorption (or emission) spectrum are here simply proportional to the FT of the autocorrelation function of the propagated wave function. 
Here, the wave functions must be propagated in real-time to access the vibronic spectra, as opposed to the previous section where propagation was done in imaginary time to access relaxed vibronic states.

\subsubsection*{Modelling absorption}
\label{sec:modelling_absorption}
For absorption and the initial wave packet, we choose the vibrational ground state (fundamental of the 3D harmonic oscillator for 81, 87, and 88) multiplied by either the diabatic state A\textsubscript{1} or the diabatic state B\textsubscript{2} as for the vibronic states relaxations. 
The choices of A\textsubscript{1} or B\textsubscript{2} has little or no influence on the shape of the absorption spectrum obtained in the end, even though they correspond to distinct polarization processes.

\begin{figure}[!htb]
    \centering
    \includegraphics[width=1\linewidth]{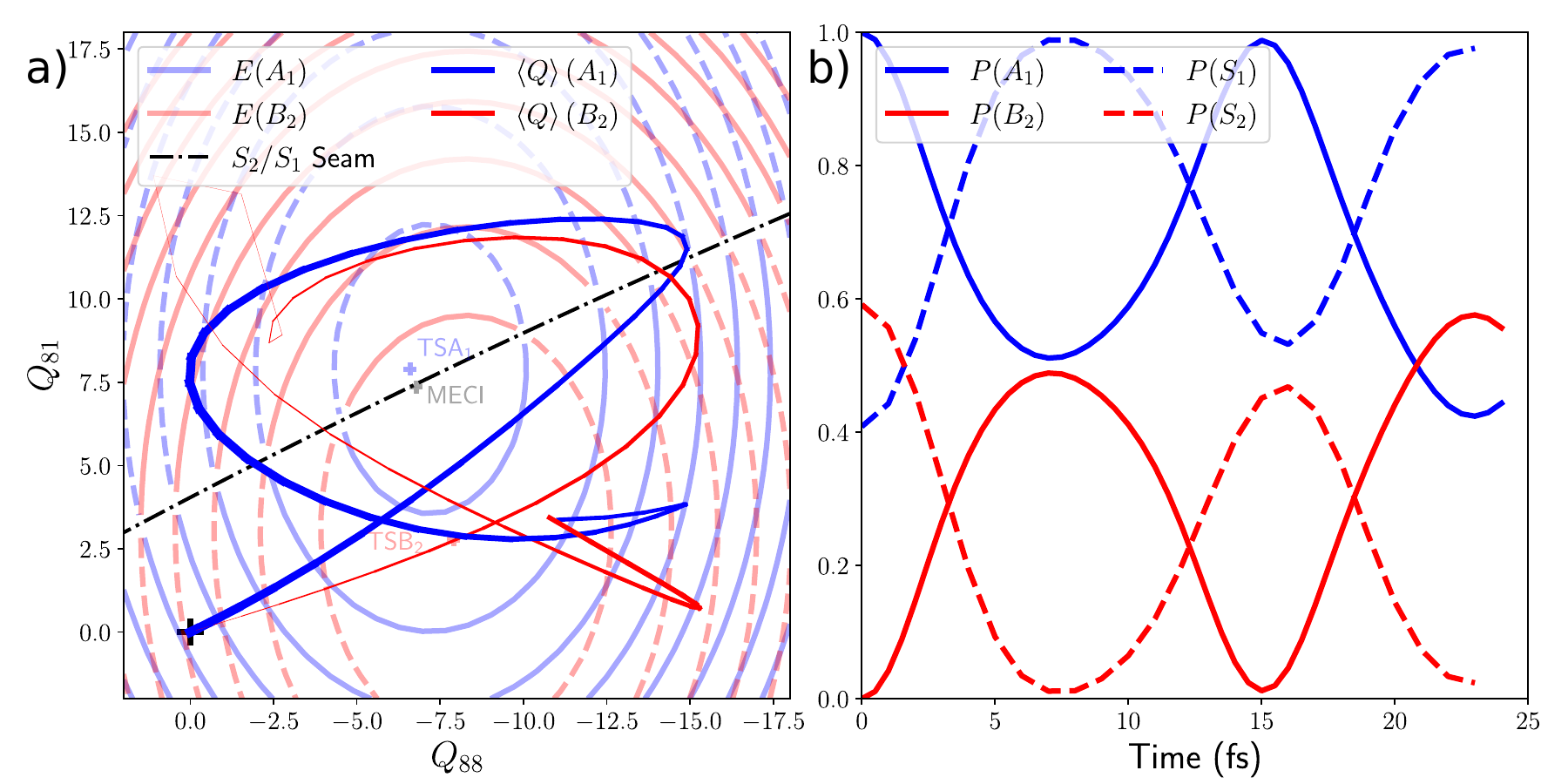}
    \caption{Early dynamics: trajectories of the center of the wave packets in the (81,88)-plane.
    Diabatic potential energy surfaces A\textsubscript{1} and B\textsubscript{2} in blue and red respectively, adiabatic PESs S\textsubscript{1} and S\textsubscript{2} in dashed lines and plain lines respectively (a). 
    Diabatic (plain lines) and adiabatic (dashed lines) populations associated to these trajectories (b).
    Propagation for an initial state on A\textsubscript{1} is given here, propagation for an initial state on B\textsubscript{2} is given in fig. SI2}
    \label{fig:early_dynamics}
\end{figure}

Before discussing the absorption spectrum, let us comment the early dynamics (first \SI{25}{\femto\second} of the \SI{200}{\femto\second} of simulation time) of the real-time propagation of the initial wave packets in the electronic excited states. 
We discuss first the propagation of the A\textsubscript{1}-populated initial state. 
The trajectories of the centers (coordinate expectation values) of the wave packets in the (81,88) plane for the two diabatic states are given in \cref{fig:early_dynamics} (a), where the PESs of the A\textsubscript{1} and B\textsubscript{2} states are shown, along with the positions of the conical intersection seam and the MECI. 
We also report diabatic and adiabatic populations (computed using the Quantics package), \cref{fig:early_dynamics} (b). 
First, let us notice that the initial state has a population of about 0.6 and 0.4 for the adiabatic states S\textsubscript{1} and S\textsubscript{2}. 
This almost 50/50 population of the first two adiabatic states is not troublesome because, in the FC region, their energy difference is small ($\Delta E = \SI{0.04}{\electronvolt}$ at our level of theory) and both are optically bright (oscillator strengths $f_1= 1.71$, $f_2 = 0.37$ in atomic units). 
Besides, in our 3D model, the ordering of the diabatic states B\textsubscript{2} and A\textsubscript{1} corresponding to adiabatic states S\textsubscript{1} and S\textsubscript{2} respectively is consistent with the \textit{ab initio} data. 
It indicates the importance of taking into account both A\textsubscript{1} modes 81 and 88 for modelling m22 as the conical intersection seam and the FC gradients are oblique within the (81,88) plane. 
Using a minimal model on the acetylenic vibrations with only modes 87 and 88 would initiate the propagation on the wrong side of the conical intersection seam, with an ordering of the diabatic states that is wrong compared to the \textit{ab initio} data. 
The multiple crossings of the conical intersection seam in the early dynamics stress the importance of having a consistent ordering of the diabatic states at the FC point and thus the importance of taking into account the coordinate 81.
On a final note, the numerical kink of the trajectory in the B\textsubscript{2} state is unconsequential because of the almost-zero diabatic population of the state B\textsubscript{2} at this time of the simulation, and seems to be of no relevance for resuming a satisfactory trajectory for times with non-zero diabatic population.
Analogous interpretations of the early dynamics can be done for the other choice of initial diabatic state, with a B\textsubscript{2}-initially populated state, for which figures are given in fig. SI2.

\begin{figure}[!htb]
    \centering
    \includegraphics[width=0.95\linewidth]{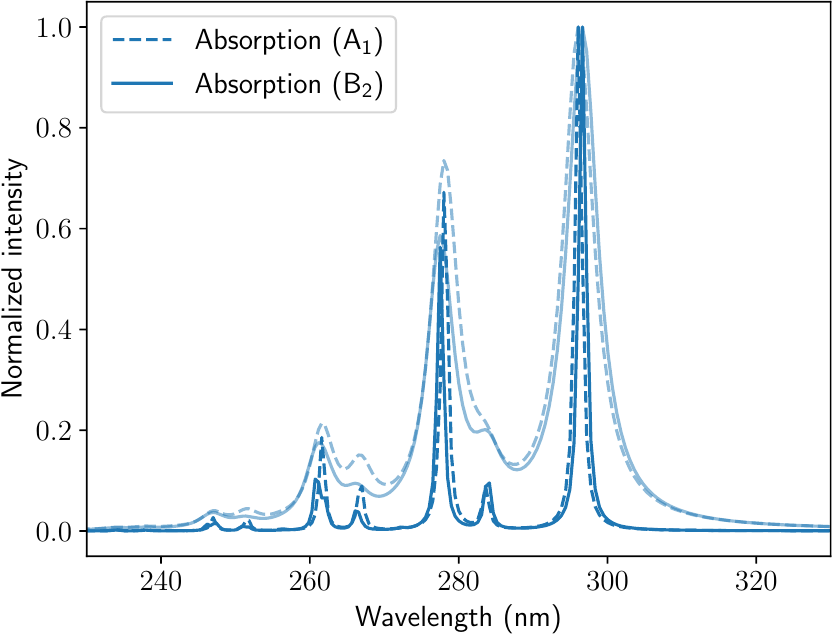}
    \caption{Simulated absorption spectra from electronic ground state to electronic excited state B\textsubscript{2} (plain lines) and A\textsubscript{1} (dashed lines).}
    \label{fig:absorption_spectra}
\end{figure}

The two propagations performed here yield two absorption spectra according to the populated excited diabatic state and corresponding light polarization. 
Let us notice that two representations are always given for each of our theoretical spectra. 
A representation obtained with FT on the autocorrelation damped with $\tau = \SI{100}{\femto\second}$ and another one one with FT on an autocorrelation function damped with $\tau = \SI{19}{\femto\second}$. 
The first representation resembles to a stick-spectrum while the second one shows realistically broadened bands. 
The damping time $\tau$ was chosen so as to match the experimental spectra. 
With our model, we access two distinct but almost identical absorption spectra (\cref{fig:absorption_spectra}). 
This is consistent with the results obtained for the vibronic states relaxation.
If analogous vibronic states are found from the two different initial states, then absorption spectra must also be analogous because the same eigenstates are accessible for the system.
Here we discard for now the differences in oscillator strength between the two diabatic states, but as both are bright, both simulations and both absorption spectra are relevant. 
Notice that within the 3D LVC model, the $0-0$ band corresponds to a transition at \SI{296}{\nano\meter}. 
The second most intense band is at \SI{278}{\nano\meter}, which corresponds to a difference of $\Delta \bar{\nu} = \SI{2190}{\per\centi\meter}$. 
The latter energy difference is expected for the acetylenic mode 87, optically active in the absorption, for which the frequency is about \SI{2200}{\per\centi\meter} in the 3D LVC model of the electronic excited states.
The other transition at \SI{284}{\nano\meter}, corresponding to $\Delta \bar{\nu} = \SI{1430}{\per\centi\meter}$ with respect to the $0-0$ band, is the expected transition involving the quinoidal mode 81.
The absorption spectra are also reported in \cref{fig:abs_emi_spectra} for completeness of the theoretical results and in \cref{fig:exp_and_theo_spectra} for comparison with the experiments.

\subsubsection*{Modelling the emission spectrum}
The emission spectrum is simulated by propagating the initial nuclear wave packet in the electronic ground state, for which the PES is described as a 3D harmonic potential with frequency parameters for mode 81, 87, and 88 extracted from \textit{ab initio} calculations at the minimum of the electronic ground state (see \cref{tab:grid}). 
For the initial state, the choice is more complicated than for simulating absorption. 
Indeed, as mentioned earlier, the first two vibronic states in the twofold excited electronic structure are quasi-degenerate and share similar characteristics in vibrational shape and populations. 
The assumption that the emission would occur only from the lowest vibronic state should not be completely valid, but we focus first on this case as the second one is analogous. 
Emitting from the lowest excited vibronic state (initially B\textsubscript{2}-populated), we identify two simple initial states for the quantum dynamics which will result in the end to two contributions to the emission spectrum. 
The two initial states are: $\chi_{A_1,0}\phi_0$ and $\chi_{B_2,0}\phi_0$. 
As illustrated in \cref{fig:vibronic_states} the first vibrational part $\chi_{A_1,0}\phi_0$ is odd with respect to coordinate 87 while the second one $\chi_{B_2,0}\phi_0$ is even with respect to coordinate 87.

\begin{figure}[!htb]
    \centering
    \includegraphics[width=1\linewidth]{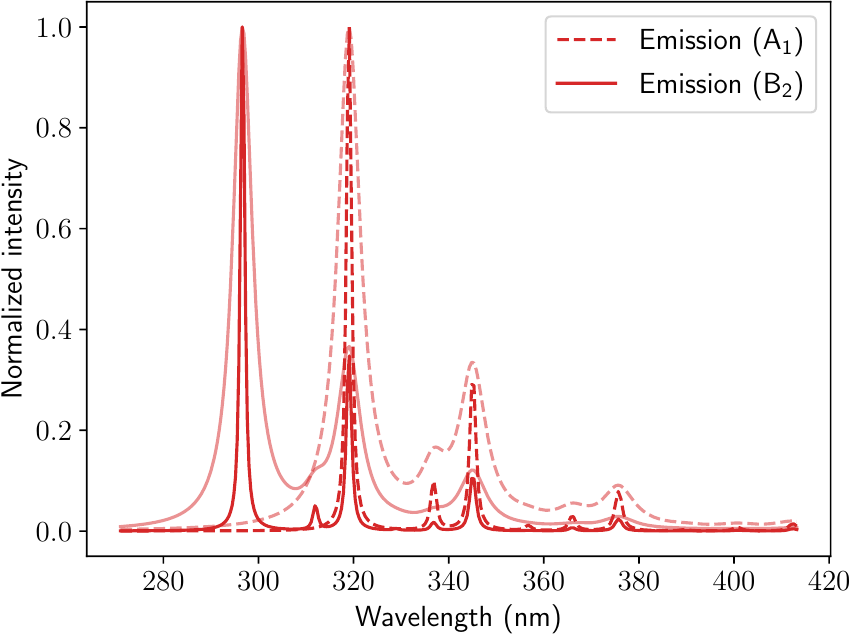}
    \caption{Simulated emission spectra from vibrational contributions of B\textsubscript{2} (plain lines) and A\textsubscript{1} (dashed lines) of the first vibronic eigenstate to the electronic ground state.
    Simulated emission spectra from the second vibronic eigenstate are given in fig. SI4.
    }
    \label{fig:emission_2a_2b}
\end{figure}

\begin{figure}[!htb]
    \centering
    \includegraphics[width=1\linewidth]{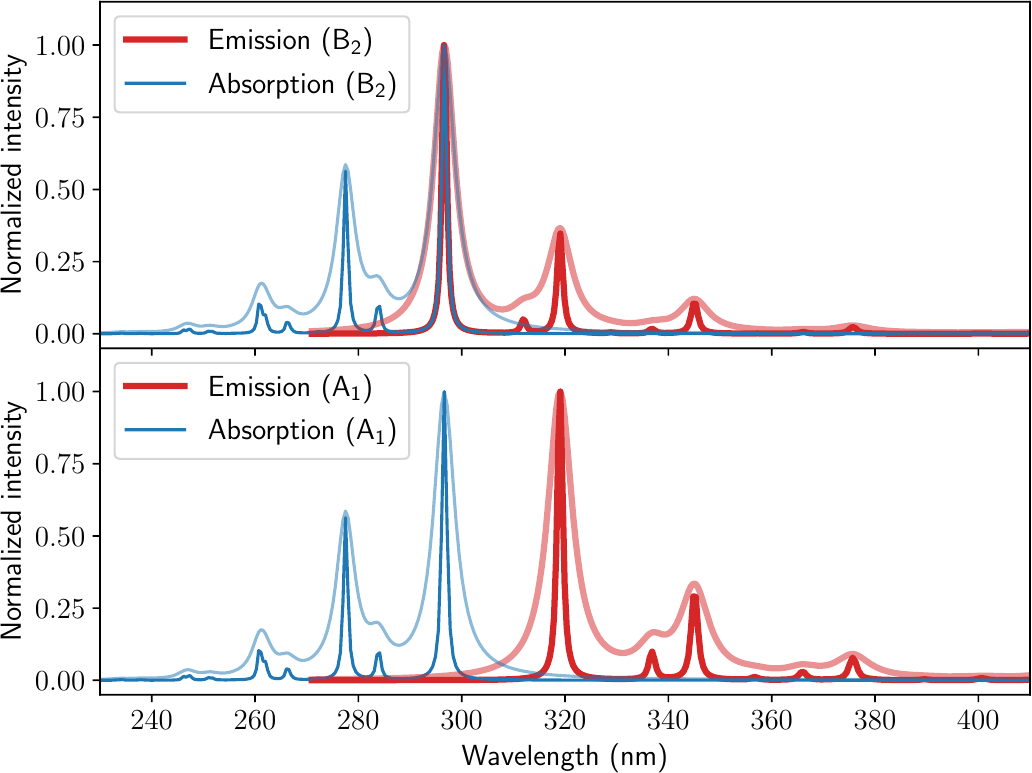}
    \caption{Absorption spectra from the electronic ground state to both diabatic states (blue lines) and emission spectra from the different contributions of the first vibronic eigenstate (red lines). 
    Simulated emission spectra from the first-excited vibronic state of the coupled S\textsubscript{1/S\textsubscript{2} manifold} are given in fig. SI5.
    Realistically broadened bands are obtained using a damping time $\tau = \SI{19}{\femto\second}$.}
    \label{fig:abs_emi_spectra}
\end{figure}

For the first vibronic eigenstate, two contributions to the emission spectrum are obtained (see \cref{fig:emission_2a_2b}).
The $0-0$ band for emission is recovered from the contribution of the vibrational part on B\textsubscript{2}, at \SI{297}{\nano\meter}. 
A vibronic progression (\SI{319}{\nano\meter} and \SI{345}{\nano\meter}) analogous to absorption is obtained as expected for frequencies in the acetylenic regions, with $\Delta \bar{\nu} = \SIlist{2320;2360}{\per\centi\meter}$. 
Such a contribution to the emission spectrum will be called a non-Stokes contribution in the following. 
The contribution of the vibrational part on A\textsubscript{1}, on the contrary, is shifted to the red wavelengths with an origin at \SI{319}{\nano\meter}, corresponding to a Stokes shift $\Delta \bar{\nu} = \SI{2320}{\per\centi\meter}$, and maintains the expected vibronic progression from this origin band to larger wavelengths. Such a contribution to the emission spectrum will be called a Stokes contribution in the following.
The positions of the first three transitions and the associated shifts with respect to the most intense band are summarized in \cref{tab:spectroscopic_data}.
\setlength{\tabcolsep}{18pt}
\begin{table*}[!hbt]
    \centering
    \begin{tabular}{l|c|c}
        \toprule
         Spectrum contribution & $\lambda_\text{max}$ (\si{\nano\meter}) & First vibronic progression (\si{\per\centi\meter}) \\
         \midrule
         Absorption & \textbf{296}, (284), 278 & --, (1430), 2190\\ 
         Emission & & \\
         \hspace{0.7cm}$\Psi_0$, non-Stokes & \textbf{297}, (312), 319 & --, (1620), 2320\\
         \hspace{0.7cm}$\Psi_0$, Stokes & \textbf{319}, (337), 345 & --, (1670), 2360\\         
         \hspace{0.7cm}$\Psi_1$, non-Stokes & \textbf{296}, (311), 318 & --, (1630), 2340\\
         \hspace{0.7cm}$\Psi_1$, Stokes & \textbf{319}, (336), 344 & --, (1590), 2280\\
         \bottomrule
    \end{tabular}
    \caption{Spectroscopic data for theoretical absorption and emission spectra, before any comparison with the experiments. 
    One row is given for absorption as both computed spectra are identical for reading the transition energies.
    The four contributions to the emission spectrum are given next.
    The most intense transition for each contribution is given in bold. 
    The second transition with low intensity is given in parenthesis.
    Associated vibronic progression are given with respect to the most intense transition of each contribution.
    }
    \label{tab:spectroscopic_data}
\end{table*}
\setlength{\tabcolsep}{6pt}
Let us notice that all these results are expected given the form of the vibrational contributions in the vibronic states. 
Indeed, with a time-independent reasoning, the vibrational part $\chi_{B_2,0}$ of the vibronic state is expected to overlap with the vibrational ground state $\chi_0$ resulting in a band origin in the same energy region than for the band origin of the absorption.
However, the vibrational part $\chi_{A_1,0}$ of the vibronic state is expected to overlap with the first vibrational excited state $\chi_1$ resulting in the shifted origin band at \SI{319}{\nano\meter}. 
The same interpretation holds for the second vibronic state, only interchanging the roles of A\textsubscript{1} and B\textsubscript{2} diabatic states.
All the transitions observed between the ground vibronic states and the excited vibronic states are schematized in the interaction diagram \cref{fig:interaction_diagram}, all the resulting contributions to the absorption and emission spectra (two for each vibronic state) are collected in \cref{fig:abs_emi_spectra}.
\subsubsection*{Discussion and comparison to the experiments}
\begin{figure}[!htb]
    \centering
    \includegraphics[width=1\linewidth]{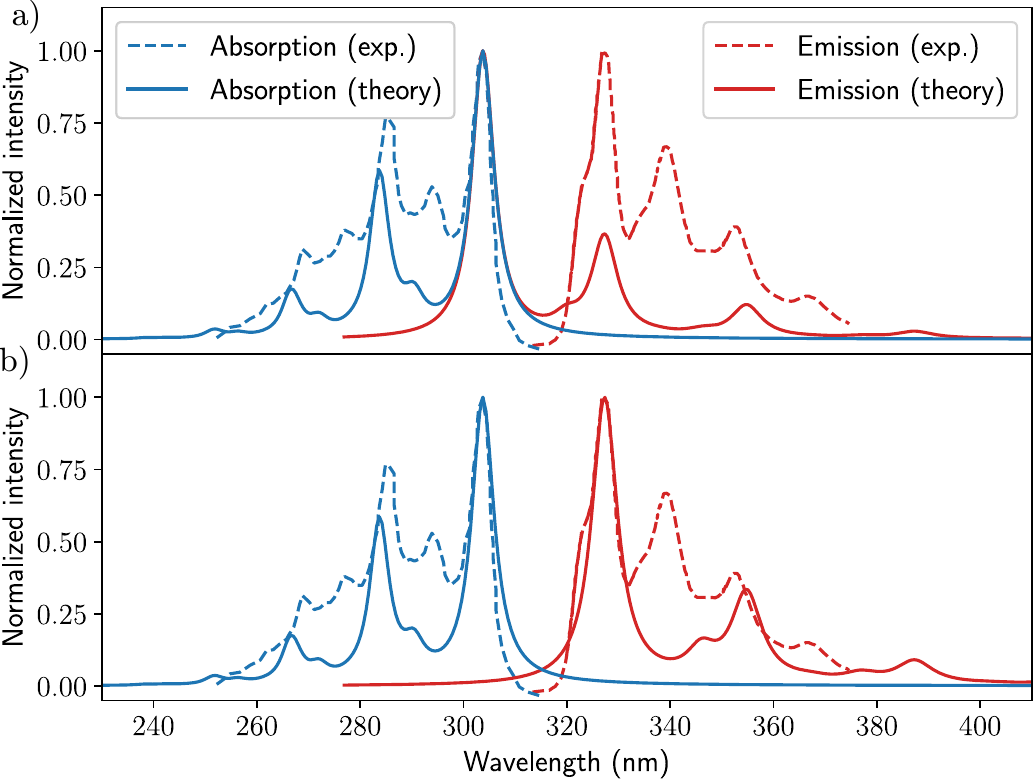}
    \caption{Experimental absorption and emission spectra (dashed lines), courtesy of Chu and Pang, 2004\cite{chu_vibronic_2004}, reproduced in (a) and (b).
    Theoretical absorption spectrum to the first vibronic eigenstate reproduced in (a) and (b), theoretical emission spectrum from the first vibronic eigenstate, contributions of B\textsubscript{2} (a) and of A\textsubscript{1} (b).
    Realistically broadened bands are obtained using a damping time $\tau = \SI{19}{\femto\second}$.
    Simulated absorption and emission spectra from the second vibronic eigenstate are given in fig. SI6.
    }
    \label{fig:exp_and_theo_spectra}
\end{figure}

We chose to compare our theoretical results to the experiments of Chu and Pang because of the exact same substitution of the molecule in both studies and because of the temperature study\cite{chu_vibronic_2004}. 
Indeed, the measurement of the spectra was done at \SIlist{25;-108;-198}{\celsius} and we choose the latter measurement for comparison.
The experimental spectra at \SI{-198}{\celsius} are reproduced in \cref{fig:exp_and_theo_spectra}, with courtesy of Chu and Pang, 2004\cite{chu_vibronic_2004}.
The theoretical spectra (considering the absorption to and emission from the first vibronic state) are superimposed to the experimental ones, shifting the $x$-axis of the theoretical spectra so that the band origins from theoretical absorption (\SI{297}{\nano\meter}) and experimental absorption (\SI{304}{\nano\meter}) match, which represents a shift of \SI{7.3}{\nano\meter} at the $0-0$ band of absorption.
The shift applied to our theoretical spectra for the $0-0$ band of absorption to match with the experiments is thus $\Delta E = \SI{0.10}{\electronvolt}$ and $\Delta \bar{\nu} = \SI{818}{\per\centi\meter}$.
This difference between the experiments and our model can be understood as an underestimated zero-point energy difference for the minimal 3D model, because other vibrations should be included to the complete description of the system. 
Besides, our model is built on data at a TD-DFT level of theory, which may imperfectly reproduce vertical transition energies. 
We stress that we shift the $x$-axis of  all our theoretical spectra of the same values of \SI{818}{\per\centi\meter} and thus do not introduce new or artificial Stokes shifts, between emission and absorption.
The broadening of the theoretical spectra was chosen so that the FWHM $\Delta \omega$ of the origin band for the absorption is the same as in the experiments. 
To ensure this, we damp the autocorrelation function with the damping time $\tau = \SI{19}{\femto\second}$.

For absorption, the vibronic progression is well reproduced with our model for transitions involving acetylenic vibrations.
The missing transitions in-between (\SIlist{294;277}{\nano\meter}) should correspond to transitions involving other vibrations, not taken into account in our minimal 3D model, such as quinoidal and triangular vibrations, expected and identified in the absorption spectra of p2 and m22 in previous work\cite{ho_vibronic_2017}. 
In the emission, the Stokes shift contribution is always identified as the contribution from the vibrational part of the vibronic states that underwent a transfer from one diabatic state to another. 
The complementary vibrational part shows no contribution to a spectrum with a significant Stokes shift. 
In the case of the Stokes contribution, the theoretical spectrum also reproduce the expected vibronic progression, with the missing bands expected to result from transitions involving other vibrations.

Strictly speaking, the spectrum to be compared to the experiments should be a combination of the four theoretical contributions described, involving the transition dipole moments for states A\textsubscript{1} and B\textsubscript{2} along with factors to take into account the population of these two states in the vibronic states. 
However, searching such a combination within this model and reasoning would never result in a predominant Stokes contribution to the emission spectrum. 
Indeed, the vibrational part of the vibronic states that overlaps with the $\chi_1$ ground vibrational state never correspond to more than 20\% of the diabatic population. 
In addition, the oscillator strength for the B\textsubscript{2} state, in the direction of $y$, is much greater than the oscillator strength of the other state (along $z$). 
As it stands, our model is still limited by several approximations on the system and preconceptions of how to simulate the phenomenon under study. 
For the reasons mentioned above, it cannot explain why the Stokes contribution dominates over the non-Stokes one, but it shows that there is a rationale for the Stokes contribution that could be based on symmetry considerations and thus may provide some tentative avenues for further explorations.

\section{Conclusions and Outlook}
In this work, we have devised diabatic linear and quadratic vibronic coupling models of Hamiltonians for the study of unusually strongly nondiabatically coupled excited states in a symmetric meta-substituted PPE (m22) and used them for spectral simulations based on quantum dynamics. 
We showed that molecular symmetry may play a major role in the unusual Stokes shift observed for this system regarding its band origin; other effects are still under study, such as the role of soft modes, solvent relaxation dynamics, and perhaps the effect of possible competing photo-isomerizations.

On practical terms, our study is expected to serve two main purposes: laying the basic foundations for a more general model of nonadiabatically coupled PESs for symmetrical or asymmetrical meta- or tri-substituted PPEs, and using it for providing a possible geometric explanation to the unusual spectroscopic properties of m22 and its derivative species.

As pointed out, our modelling of m22 suggests some plausible rationale (based on molecular vibronic symmetry and light polarization transfer or conservation) for distinct different Stokes and non-Stokes contributions to the emission spectrum, with no fully conclusive elucidation yet on why the Stokes contributions should be predominant, as indeed observed in the experiments.

It must be understood that modelling emission (as opposed to absorption) spectra is not a routine task and potentially raises other fundamental questions about how we should consider the very role and nature of the stationary states involved in the experimental acquisition of absorption and emission spectra against the duration of the excitation light source and the time for the environment (external: solvent; internal: soft modes) to fully relax compared to the time it takes for an excited molecule to emit light.

Within our description, for the Stokes contribution to overcome the non-Stokes contribution, the effect of the coupling would have to be stronger, hence fully forcing the electronic population to be affected quantitatively by the non-totally symmetrical coordinate 87 and its odd character (according to the linear behavior of the coupling with respect to it), along the way of a geometric (Berry or Longuet-Higgins) phase effect such that the nuclear density goes to the other side of the conical intersection with a nodal line.

This is a possibility, but it obviously is very sensitive to the level of theory used for producing \textit{ab initio} data. However, in the extreme case of a completely transferred population to a non-totally symmetrical wave packet, the $0-0$ band in the absorption spectrum should also disappear, which thus seems to raise some contradiction.

Such hypotheses are thus to be evaluated carefully in future works, considering for example additional but relevant degrees of freedom for the vibronic transitions observed experimentally in the steady-state spectra, or dissipation effects (for example in the context of the hierarchical equations of motion that we recently explored together with our colleagues on the same system but within another context)\cite{jaouadi_laser-controlled_2022}. 
More extensive simulations are currently under study and will be addressed soon in future work.

Finally, it must be noted as a last hypothesis that the non-Stokes contributions to the emission identified in this work seem to be associated to some unconventionally ultrafast coherently polarized resonant Raman process, while usual fluorescence spectra are typically viewed as incoherent, slow, and spontaneous, and they should exhibit both Stokes and non-Stokes contributions.
Such a mystery certainly now awaits for further detailed comparisons with time-resolved spectroscopic experimental studies.

\section*{Supplementary Material}
Supplementary material is available for this work. 
It includes a complete description of the LVC and QVC parameters along with the associated operator files for the \texttt{Quantics} package and the contour plots of the adiabatic QVC PESs.
Complementary figures of the first-excited vibronic state of the coupled excited-state manifold and the emission spectra contributions from it are also given. 

\section*{Acknowledgments}
The authors warmly thank Y. Pang and Q. Chu for allowing the reproduction of the experimental absorption and emission spectra of 1,3-bis(phenylethynyl)benzene (m22).
The authors would also like to personally thank M. Desouter-Lecomte and E. Mangaud for very fruitful discussions on the system and the emission phenomenon, and their incentive for plausible extensions to further studies with complementary viewpoints. 
J. G. acknowledges the French MESR (Ministère de l’Enseignement supérieur, de la Recherche) and the ENS (Ecole Normale Supérieure) of Lyon for funding his PhD grant, hosted at the University of Montpellier.
\section{Appendices}
\label{sec:appendices}
\subsection{Duschinsky matrix and shift vectors}
\label{sec:duschinsky}
Normal modes of vibration are computed at the minima of the ground electronic state and the first excited electronic state. The transformation from the normal modes of vibration at the minimum of the first electronic excited state, $\boldsymbol{Q}'$ to the normal modes of vibration at the minimum of the electronic ground state $\boldsymbol{Q}''$ is as follows:
\begin{equation}
\boldsymbol{Q}'=\boldsymbol{J}\boldsymbol{Q}'' + \boldsymbol{K}
\end{equation}
where $\boldsymbol{J}$ and $\boldsymbol{K}$ are the so-called Duschinsky matrix and shift vector, respectively.
For modes 81, 87, and 88, the Duschinsky matrix is:
\begin{equation}
\begin{bmatrix}
-0.118694 & -0.000010 & 0.000187\\
-0.050807 & 0.733369 & 0.662382\\
-0.005732 & -0.671423 & 0.740842
\end{bmatrix}
\end{equation}
and the shift vector is
\begin{equation}
\begin{bmatrix}
-4.01414\\
7.25456\\
7.59322
\end{bmatrix}.
\end{equation}
The Duschinsky matrix characterises the rotation from the normal modes of the C\textsubscript{2v} S\textsubscript{0} minimum to those of the C\textsubscript{s} S\textsubscript{1} minimum within an adiabatic approximation. 
An illustration is given in \cref{fig:axes}, where we see that the S\textsubscript{0} delocalised stretching modes 87 (B\textsubscript{2}) and 88 (A\textsubscript{1}) fully mix so as to bring stretching modes that are localised on the left or on the right branch (one or the other p2 pseudofragment within m22). 
The shift vector consistently expands over A\textsubscript{1} and B\textsubscript{2} modes so as to go from the S\textsubscript{0} minimum to one or the other S\textsubscript{1} minima. C\textsubscript{2v} symmetry gets broken from the S\textsubscript{0} minimum to the S\textsubscript{1} minima, such that the matrix is full. 
We can thus distinguish between primary and secondary effects. 
As long as C\textsubscript{2v} symmetry is preserved, ($Q_{87}=0$), both adiabatic and diabatic representations coincide and Hessians exhibit a block-diagonal structure with respect to A\textsubscript{1} and B\textsubscript{2} modes (primary Duschinsky effect between 81 and 88; $\gamma$-terms). 
As soon $Q_{87}$ varies, the symmetry is broken from C\textsubscript{2v} to C\textsubscript{s}, and Hessians are then “full” matrices (secondary Duschinsky effect between 81 and 87 or 88 and 87; $\mu$-terms).

Note that the correlation between modes 81 and 87 or 88 is negligible (less than 0.3\% in any case). 
These small correlations suggest the validity of neglecting the primary Duschinsky parameters, $\gamma_{81,88}^{(k)}$ in the QVC model.
The correlation between 87 and 88 corresponds to a \SI{45}{\degree} rotation of the ground state normal modes to obtain the excited state normal modes, which is consistent with the results found for the orientation of the ellipsoids in the plane (87,88) of the QVC model PESs.

\end{document}



\title[SI: Unusual Stokes shift and vibronic symmetry]{\emph{Supplementary Information to:}\\On the unusual Stokes shift in the smallest PPE dendrimer building block:\\Role of the vibronic symmetry on the band origin?}



\author{Joachim Galiana}
\author{Benjamin Lasorne}%
\affiliation{ 
ICGM, Univ Montpellier, CNRS, ENSCM, Montpellier, France
}
\email{benjamin.lasorne@umontpellier.fr}
%


\date{\today}
\maketitle 
\section{Linear and Quadratic vibronic coupling Hamiltonian models}
\subsection{Parameters}
\begin{table}[H]
\caption{LVC model parameters obtained upon fitting \textit{ab initio} calculations. 
Parameters associated to the first-order expansion ($\lambda_{i}$ or $\kappa_{i}$) are given by the associated characteristic shift $d_{i}=\pm\frac{\lambda_{i}}{k_{i}}\text{ or }-\frac{\kappa_{i}}{k_{i}}^{(k)}$ in the mass-weighted framework for displacement along the normal modes directions. a. u. stands for atomic units.
Superscripts (1) and (2) refer to electronic excited states A\textsubscript{1 and B\textsubscript{2}, respectively, coincident to S\textsubscript{2} and S\textsubscript{1} at the FC point.}}
\label{tab:parameters_LVC}
\centering
\begin{tabular}{cc|ccc}
\toprule
Parameter & Value (a.u.) & Equivalent & Value & Unit\\
  \midrule
  $E^{(1)}$ & -846.0365041588223 & -- & 4.405 & \si{\electronvolt}\\
  $E^{(2)}$ & -846.0374188254349 & -- & 4.380 & \si{\electronvolt}\\
  \midrule
  $\kappa_{81}^{(1)}$ & -0.000376319797672 & ${d_{81}^{(1)}}$  & 7.900 & $a_{0}\sqrt{m_{e}}$ \\
  $\kappa_{81}^{(2)}$ & -0.000154827457621 & ${d_{81}^{(2)}}$ & 3.094 & $a_{0}\sqrt{m_{e}}$ \\
  $\kappa_{88}^{(1)}$ & 0.000773557607664 & ${d_{88}^{(1)}}$ & -7.205 & $a_{0}\sqrt{m_{e}}$ \\
  $\kappa_{88}^{(2)}$ & 0.000894008947134 & ${d_{88}^{(2)}}$ & -8.274 & $a_{0}\sqrt{m_{e}}$ \\
  $\lambda_{87}$ & 0.000745662804364 & $d_{87}$ & 7.417 & $a_{0}\sqrt{m_{e}}$ \\
  \midrule
  $k_{81}^{(1)}$ & 0.0000476374261878 & $\omega_{81}^{(1)}$& 1515 & \si{\per\centi\meter}\\
  $k_{81}^{(2)}$ & 0.0000500377681987 & $\omega_{81}^{(2)}$& 1552 & \si{\per\centi\meter}\\
  $k_{88}^{(1)}$ & 0.000107366759493 & $\omega_{88}^{(1)}$& 2274 & \si{\per\centi\meter}\\
  $k_{88}^{(2)}$ & 0.000108052889481 & $\omega_{88}^{(2)}$& 2281 & \si{\per\centi\meter}\\
  $k_{87}^{(1)}$ & 0.000100509641682 & $\omega_{87}^{(1)}$& 2200 & \si{\per\centi\meter}\\
  $k_{87}^{(2)}$ & 0.000100546482724 & $\omega_{87}^{(2)}$& 2201 & \si{\per\centi\meter}\\
  \bottomrule
\end{tabular}
\end{table}
\begin{table}[H]
\caption{QVC model parameters obtained upon fitting \textit{ab initio} calculations. 
Parameters associated to the first-order expansion ($\lambda_{i}$ or $\kappa_{i}$) are given by the associated characteristic shift $d_{i}=\pm\frac{\lambda_{i}}{k_{i}}\text{ or }-\frac{\kappa_{i}}{k_{i}}^{(k)}$ in the mass-weighted framework for displacement along the normal modes directions. a. u. stands for atomic units.
Parameters associated to the second-order expansion ($\gamma_{i,j}$ or $\mu_{i,j}$) are only given in atomic units.
Superscripts (1) and (2) refer to electronic excited states A\textsubscript{1 and B\textsubscript{2}, respectively, coincident to S\textsubscript{2} and S\textsubscript{1} at the FC point.}}
\label{tab:parameters_QVC}
\centering
\begin{tabular}{cc|ccc}
\toprule
Parameter & Value (a.u.) & Equivalent & Value & Unit\\
  \midrule
  $E^{(1)}$ & -846.0364194742663 & -- & 4.407 & \si{\electronvolt}\\
  $E^{(2)}$ & -846.0374459505783 & -- & 4.379 & \si{\electronvolt}\\
  \midrule
  $\kappa_{81}^{(1)}$ & -0.000389737234515 & ${d_{81}^{(1)}}$  & 8.203 & $a_{0}\sqrt{m_{e}}$ \\
  $\kappa_{81}^{(2)}$ & -0.00015231303099 & ${d_{81}^{(2)}}$ & 3.059 & $a_{0}\sqrt{m_{e}}$ \\
  $\kappa_{88}^{(1)}$ & 0.000790813573696 & ${d_{88}^{(1)}}$ & -7.365 & $a_{0}\sqrt{m_{e}}$ \\
  $\kappa_{88}^{(2)}$ & 0.000894785951449 & ${d_{88}^{(2)}}$ & -8.289 & $a_{0}\sqrt{m_{e}}$ \\
  $\lambda_{87}$ & 0.00065515654139 & $d_{87}$ & 6.569 & $a_{0}\sqrt{m_{e}}$ \\
  \midrule
  $\gamma_{81,88}^{(1)}$ & -0.00000259362606264 & -- & \num{-2.594e-6} & $E_h/(a_0^2m_e)$ \\
  $\gamma_{81,88}^{(2)}$ & -0.000000511909024135 & -- & \num{-0.512e-6} & $E_h/(a_0^2m_e)$ \\
  $\mu_{87,81}$ & -0.00000108420351464 & -- & \num{-1.084e-6} & $E_h/(a_0^2m_e)$ \\
  $\mu_{87,88}$ & -0.0000118672374859 & -- & \num{-11.867e-6} & $E_h/(a_0^2m_e)$ \\
  \midrule
  $k_{81}^{(1)}$ & 0.0000475127642257 & $\omega_{81}^{(1)}$& 1513 & \si{\per\centi\meter}\\
  $k_{81}^{(2)}$ & 0.0000498104486865 & $\omega_{81}^{(2)}$& 1549 & \si{\per\centi\meter}\\
  $k_{88}^{(1)}$ & 0.000107374006285 & $\omega_{88}^{(1)}$& 2274 & \si{\per\centi\meter}\\
  $k_{88}^{(2)}$ & 0.00010794607905 & $\omega_{88}^{(2)}$& 2280 & \si{\per\centi\meter}\\
  $k_{87}^{(1)}$ & 0.000101563963056 & $\omega_{87}^{(1)}$& 2212 & \si{\per\centi\meter}\\
  $k_{87}^{(2)}$ & 0.0000979165892662 & $\omega_{87}^{(2)}$& 2171 & \si{\per\centi\meter}\\
  \bottomrule
\end{tabular}
\end{table}
\subsection{Potential energy surfaces}
\begin{figure}[H]
\centering
\includegraphics[width=1.0\linewidth]{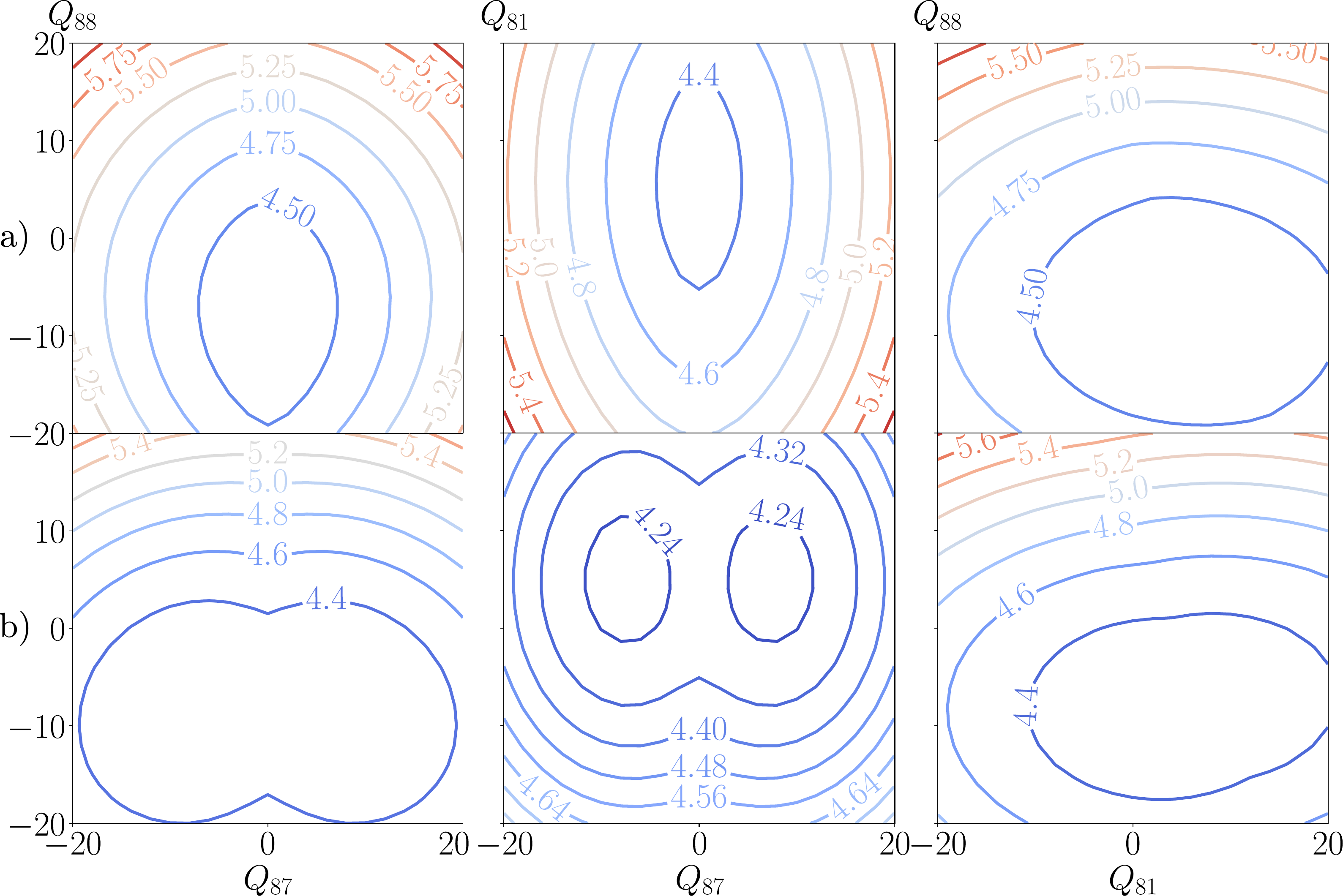}
\caption{QVC potential energy surfaces (eigenvalues of the QVC Hamiltonian model in \si{eV}) of electronic states S\textsubscript{1} (a) and S\textsubscript{2} (b) in the planes (87,81), (87,88), and (81,88) from left to right. 
For unspecified coordinates, the values are those at the MECI.
All coordinates are mass-weighted and given in atomic units, $\sqrt{m_e}a_0$.}
\label{fig:contours_all_QVC}
\end{figure}
\section{Results of quantum dynamics}
\subsection{Early dynamics with different initial conditions}
\begin{figure}[H]
    \centering
    \includegraphics[width=1\linewidth]{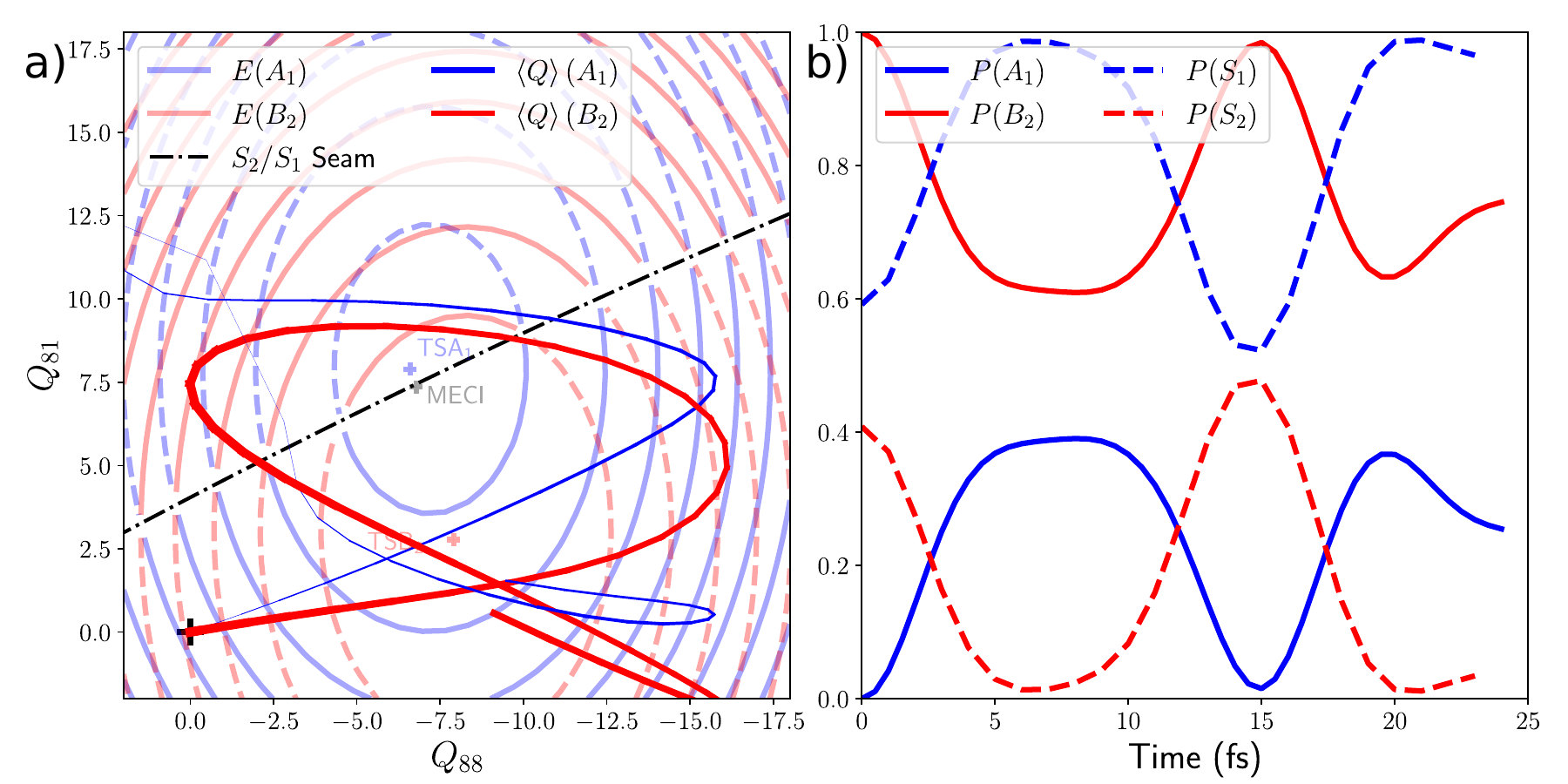}
    \caption{Early dynamics: trajectories of the center of the wave packets in the (81,88)-plane.
    The initial state for the propagation is entirely B\textsubscript{2}-populated.
    Diabatic potential energy surfaces A\textsubscript{1} and B\textsubscript{2} in blue and red respectively, adiabatic PESs S\textsubscript{1} and S\textsubscript{2} in dashed lines and plain lines respectively (a). 
    Diabatic (plain lines) and adiabatic (dashed lines) populations associated to these trajectories (b).
    }
    \label{fig:early_dynamics_B2}
\end{figure}
\subsection{Relaxation and emission from the second vibronic eigenstate}
\begin{figure}[H]
\centering
\includegraphics[width=1\linewidth]{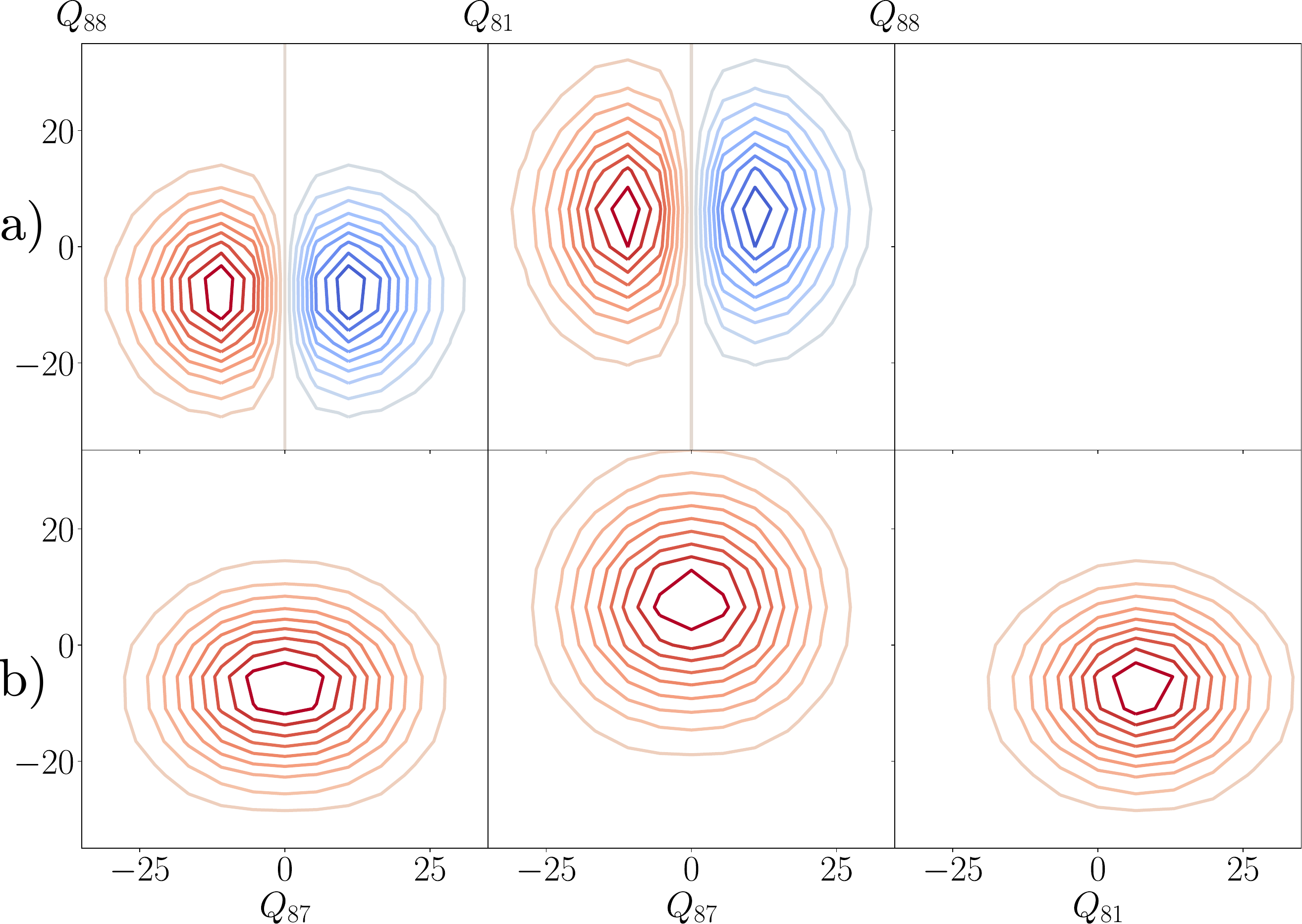}
\caption{Nuclear wave packet contributions in diabatic electronic states B\textsubscript{2} (a) and A\textsubscript{1} (b) in the first-excited vibronic state of the coupled S\textsubscript{1/S\textsubscript{2} manifold}, in planes (87,81), (87,88), and (81,88) from left to right. 
For unspecified coordinates, the values are those at the MECI.
All coordinates are mass-weighted and given in atomic units, $\sqrt{m_e}a_0$.
The blank figure corresponds to the wave functions having the B\textsubscript{2} coordinate 87 in the nodal plane.}
\label{fig:vibronic_states_2}
\end{figure}

\begin{figure}[H]
    \centering
    \includegraphics[width=1\linewidth]{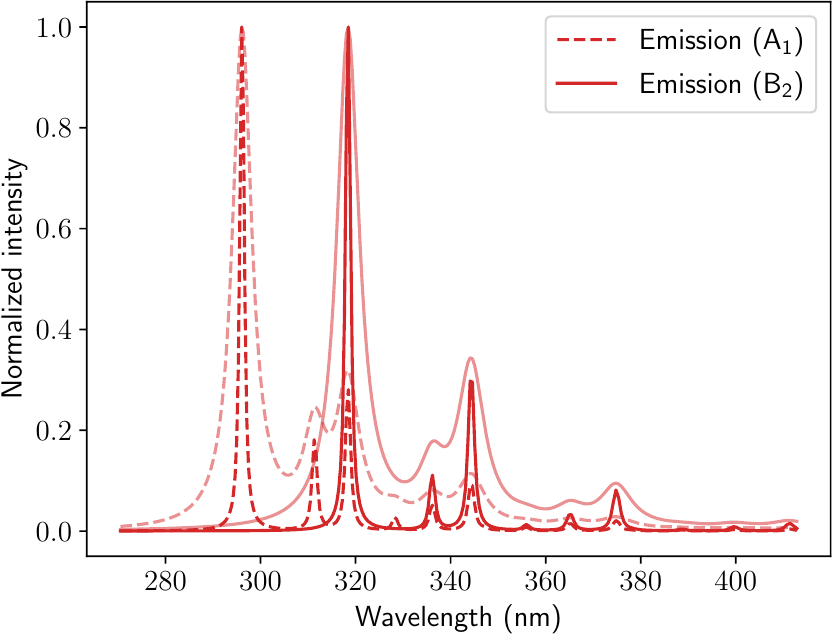}
    \caption{Simulated emission spectra from vibrational contributions of B\textsubscript{2} (plain lines) and A\textsubscript{1} (dashed lines) of the second vibronic eigenstate to the electronic ground state.
    }
    \label{fig:emission_1a_1b}
\end{figure}

\begin{figure}[H]
    \centering
    \includegraphics[width=1\linewidth]{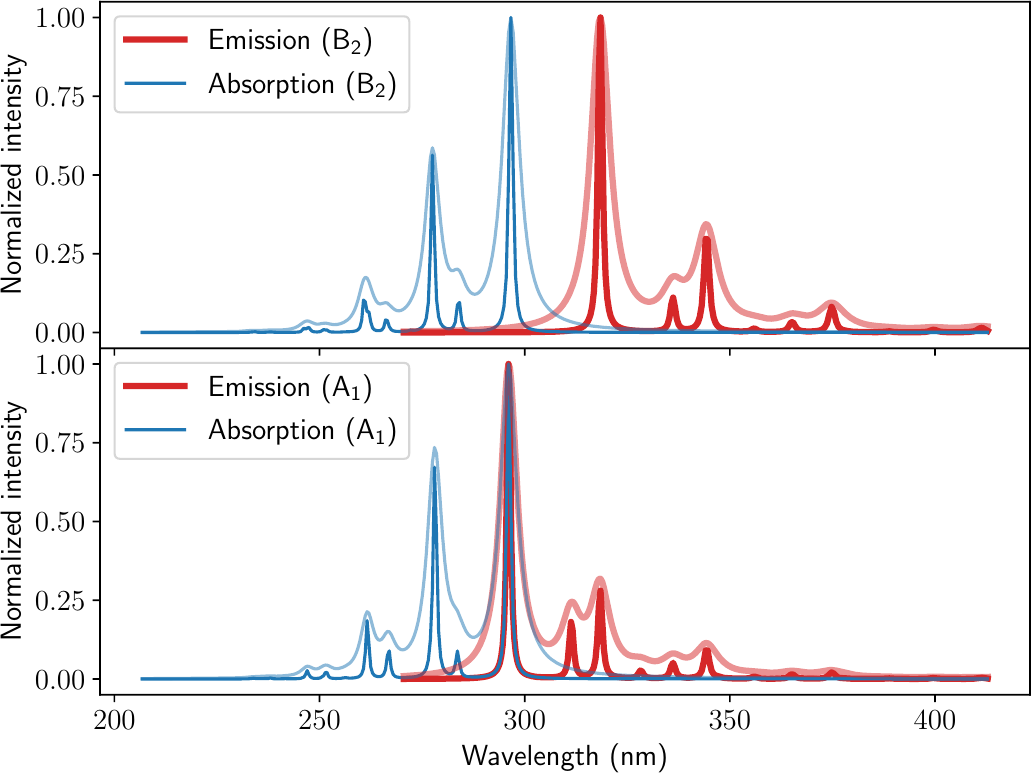}
    \caption{Absorption spectra from the electronic ground state to both diabatic states (blue lines) and emission spectra from the different contributions of the second vibronic eigenstate (red lines). 
    Realistically broadened bands are obtained using a damping time $\tau = \SI{19}{\femto\second}$.}
    \label{fig:abs_emi_spectra_1}
\end{figure}

For the comparison of the results corresponding to the second vibronic eigenstate and the experiment, the shift applied to our theoretical spectra for the 0-0 band of absorption to match with the experiment is $\Delta E = \SI{0.11}{\electronvolt}$ and $\Delta \bar{\nu} = \SI{883}{\per\centi\meter}$ (\SI{7.9}{\nano\meter} at the 0-0 band).
\begin{figure}[H]
    \centering
    \includegraphics[width=1\linewidth]{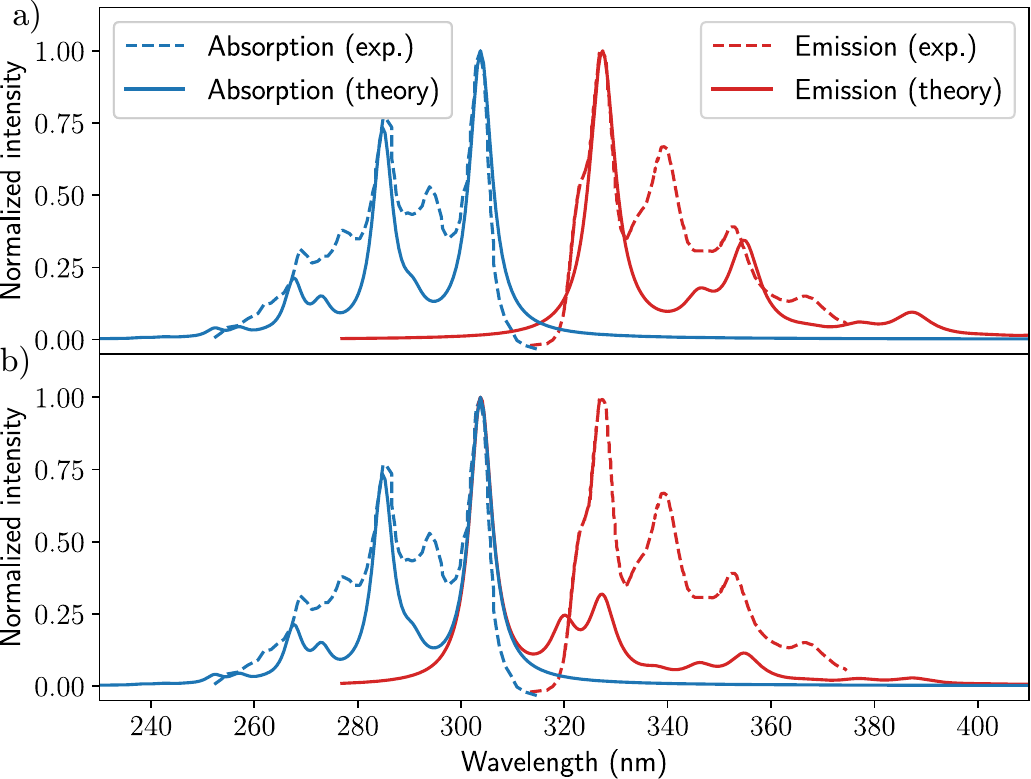}
    \caption{Experimental absorption and emission spectra (dashed lines in (a) and (b)). Data reproduced with permission from Spectrochim. Acta A 60, 1459 (2004)\cite{chu_vibronic_2004}. Copyright 2003 Elsevier B.V. 
    Theoretical absorption spectrum to the second vibronic eigenstate reproduced in (a) and (b), theoretical emission spectrum from the first vibronic eigenstate, contributions of B\textsubscript{2} (a) and of A\textsubscript{1} (b).
    Realistically broadened bands are obtained using a damping time $\tau = \SI{19}{\femto\second}$.
    }
    \label{fig:exp_and_theo_spectra}
\end{figure}
\section{Vibronic coupling operator files}
The operator files to be used with Quantics are given in the following for both LVC and QVC Hamiltonian models.
Note that the parameters are given for a mass-weighted system of coordinates.
As a consequence, the reduced masses of the three coordinate are 1, numerically.
Inputs files for propagation and relaxation calculations are available upon request.
\subsection{Quantics Operator file for the LVC Hamiltonian model} 
\noindent
\includegraphics[page=1]{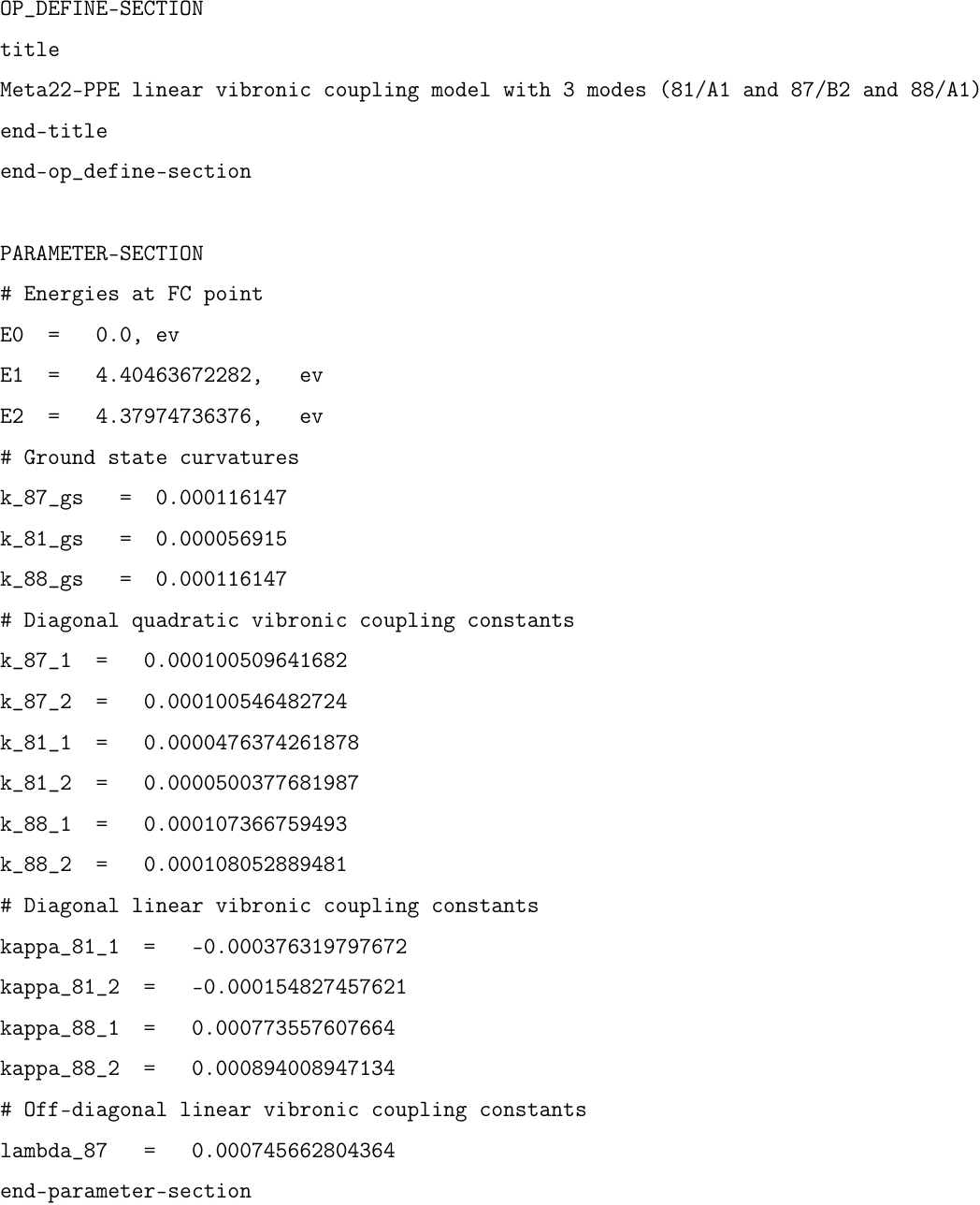}
\newpage
\noindent
\includegraphics[page=2]{operator_file_LVC-crop.pdf}
\newpage
\noindent
\includegraphics[page=3]{operator_file_LVC-crop.pdf}
\subsection{Quantics Operator file for the QVC Hamiltonian model} 
\noindent
\includegraphics[page=1]{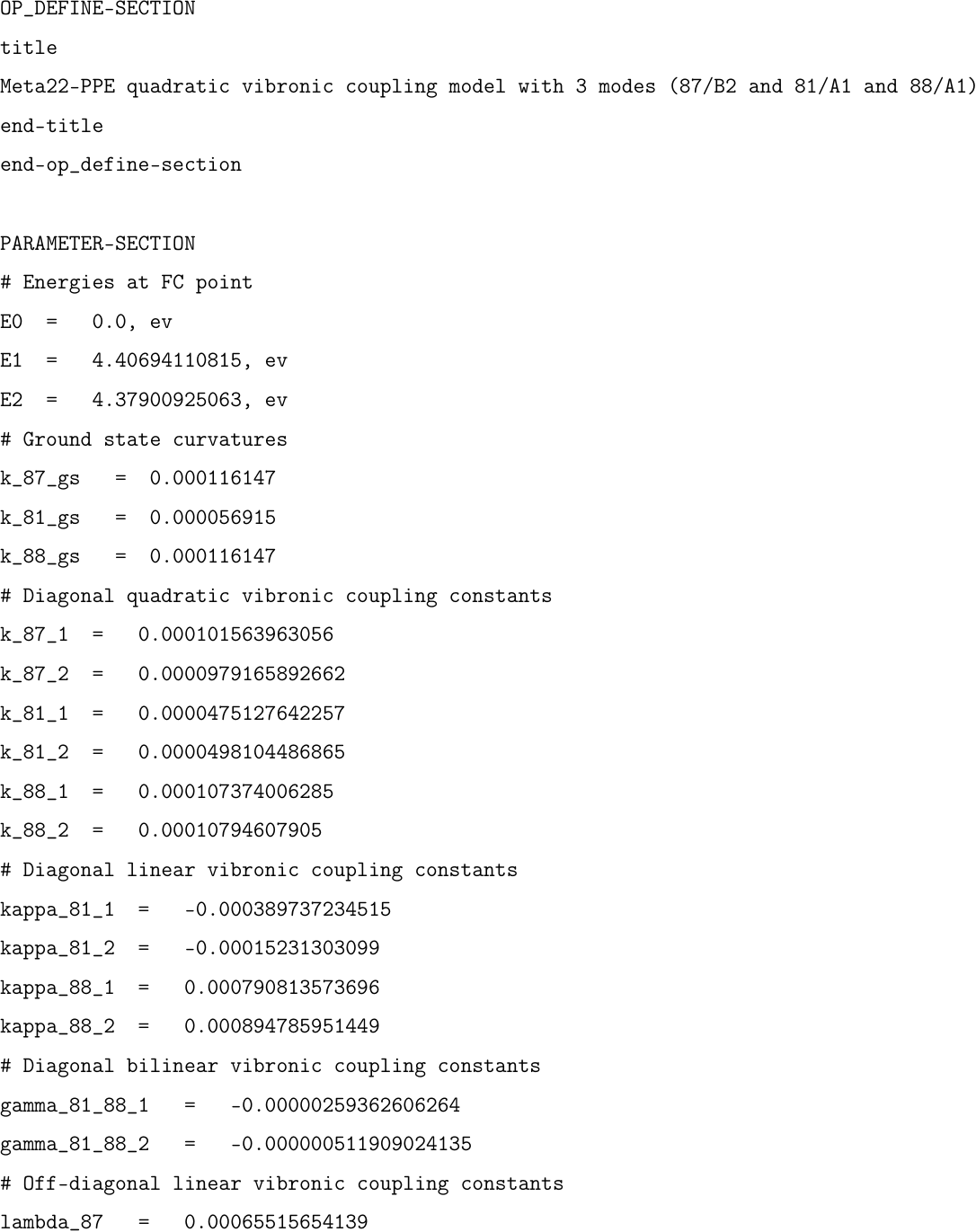}
\newpage
\noindent
\includegraphics[page=2]{operator_file_QVC-crop.pdf}
\newpage
\noindent
\includegraphics[page=3]{operator_file_QVC-crop.pdf}